\documentclass[trackchanges,twocolumn]{aastex62}
\usepackage{amsmath,amstext}
\usepackage{units}
\usepackage{url}
\usepackage{xspace}

\newcommand{\E}[1]{\times10^{#1}}
\newcommand{\msol}{ \, M_\sun}

\newcommand{\bi}{\begin{itemize}}
\newcommand{\ei}{\end{itemize}}
\newcommand{\commentOut}[1]{}
\newcommand{\sedona}{\texttt{Sedona}\xspace}
\newcommand{\flash}{\texttt{FLASH}\xspace}

\shortauthors{Shen et al.}

\begin{document}

\title{\bf \Large{Multi-Dimensional Radiative Transfer Calculations of Double Detonations of Sub-Chandrasekhar-Mass White Dwarfs}}

\author[0000-0002-9632-6106]{Ken J.\ Shen}
\affiliation{Department of Astronomy and Theoretical Astrophysics Center, University of California, Berkeley, CA 94720, USA}

\author[0000-0002-1184-0692]{Samuel J.\ Boos}
\affiliation{Department of Physics \& Astronomy, University of Alabama, Tuscaloosa, AL, USA}

\author[0000-0002-9538-5948]{Dean M.\ Townsley}
\affiliation{Department of Physics \& Astronomy, University of Alabama, Tuscaloosa, AL, USA}

\author{Daniel Kasen}
\affiliation{Department of Astronomy and Theoretical Astrophysics Center, University of California, Berkeley, CA 94720, USA}
\affiliation{Department of Physics, University of California, Berkeley, CA 94720, USA}
\affiliation{Lawrence Berkeley National Laboratory, Berkeley, CA, USA}

\correspondingauthor{Ken J. Shen}
\email{kenshen@astro.berkeley.edu}

\begin{abstract}

Study of the double detonation Type Ia supernova scenario, in which a helium shell detonation triggers a carbon core detonation in a sub-Chandrasekhar-mass white dwarf, has experienced a resurgence in the past decade.  New evolutionary scenarios and a better understanding of which nuclear reactions are essential have allowed for successful explosions in white dwarfs with much thinner helium shells than in the original, decades-old incarnation of the double detonation scenario.  In this paper, we present the first suite of light curves and spectra from  multi-dimensional radiative transfer calculations of thin-shell double detonation models, exploring a range of white dwarf and helium shell masses.  We find broad agreement with the observed light curves and spectra of non-peculiar Type Ia supernovae, from subluminous to overluminous subtypes, providing evidence that double detonations of sub-Chandrasekhar-mass white dwarfs produce the bulk of observed Type Ia supernovae.  Some discrepancies in spectral velocities and colors persist, but these may be brought into agreement by future calculations that include more accurate initial conditions and radiation transport physics.

\end{abstract}


\section{Introduction}

The identity of Type Ia supernova (SN~Ia) progenitors remains uncertain.  There is general consensus that these explosions are powered by the radioactive decay of $^{56}$Ni \citep{pank62a,cm69} produced in the explosion of a white dwarf (WD), but the mechanism of the explosion and the nature of the companion(s) are still debated (for a review, see \citealt{maoz14a}).

A double detonation of a sub-Chandrasekhar-mass WD, in which a helium shell detonation triggers a carbon core detonation, was one of the first proposed SN~Ia mechanisms \citep[e.g.,][]{nomo82b,wtw86}.  However, the model fell out of favor due to discrepancies with observed SNe~Ia imparted by the thermonuclear ash from the relatively massive ($\sim 0.1 \msol$) helium shells that arise when the donor is a low-mass, non-degenerate helium star  \citep{hk96,nuge97}.

More recently, there has been a resurgence in research into this SN~Ia explosion mechanism due in part to the realization that the helium shells at the onset of the supernova in double WD binary progenitors are orders of magnitude smaller than previously considered \citep{bild07,guil10,dan12,rask12}, yet are still massive enough to support successful shell and subsequent core detonations \citep{tmb12,moor13a,shen14a,shen14b}.  These theoretical studies have been further bolstered by the observational confirmation that double WD binaries definitively lead to sub-Chandrasekhar-mass WD explosions that are likely SNe~Ia \citep{shen18b}.  The combination of theoretical and observational motivation has led to a large number of explosion and radiative transfer simulations of sub-Chandrasekhar-mass WD detonations, with a range of physical configurations, dimensionality, nuclear reaction network complexity, and radiation transport approximations \citep[e.g.,][]{fhr07,fink10,krom10,pakm10,pakm11,pakm12b,pakm13a,pakm21a,sim10,sim12,kush13a,moll13a,moll14a,rask14a,blon17a,shen18a,shen21a,tani18b,tani19a,mile19a,poli19a,poli21a,town19a,leun20a,gron20a,gron21a,boos21a,mage21a}.  Of these, \cite{town19a} was the first study of a multi-dimensional explosion simulation utilizing a large enough reaction network to allow for a very thin ($0.02 \msol$ on a $1.0 \msol$ core) helium shell detonation.  \cite{town19a}'s work was subsequently expanded upon by \cite{boos21a}'s study of a range of thin-shell explosion models.

In this paper, we follow-up \cite{boos21a}'s work and present the first suite of light curves and spectra from multi-dimensional radiative transfer calculations of  thin-helium-shell double detonation simulations, as well as of several models with  thick helium shells for comparison to previous work.  In Section \ref{sec:mod}, we describe the starting models and \sedona, the radiation transport code we use to produce light curves and spectra, which we show in Section \ref{sec:results}.  Correlations among photometric and spectral indicators are discussed in Section \ref{sec:maxcorr}.  We compare to previous work in Section \ref{sec:prev} and summarize our results in Section \ref{sec:conc}.


\section{Description of models and calculations}
\label{sec:mod}

\begin{table*}
\begin{center}
\caption{Explosion model parameters}
\begin{tabular}{ccccc}
Total mass [$M_\odot$] & $\rho_5$\tablenotemark{a} & Core mass [$M_\odot$] & Shell mass [$M_\odot$] & Total $^{56}$Ni mass [$M_\odot$] \\
\tableline
0.85 & 2 & 0.82 & 0.033 & 0.11 \\
0.85 & 3 & 0.80 & 0.049 &  0.14 \\
1.00 & 2 & 0.98 & 0.016 &0.50  \\
1.00 & 3 & 0.98 & 0.021 & 0.51 \\
1.00 & 6 & 0.96 & 0.042 & 0.52 \\
1.00 & 14 & 0.90 & 0.100 & 0.56 \\
1.02 & 2 & 1.00 & 0.021 & 0.53 \\
1.02\tablenotemark{b} & 2 & 1.00 & 0.021 & 0.53 \\
1.10 & 2 & 1.09 & 0.0084 & 0.75 \\
1.10 & 3 & 1.09 & 0.011 & 0.76
\end{tabular}
\end{center}
\tablenotetext{a}{Shell base density in units of $\unit[10^5]{\rm g \, cm^{-3}}$.}
\tablenotetext{b}{Reduced oxygen abundance in the shell.}
\label{tab:models}
\end{table*}

We use the two-dimensional sub-Chandrasekhar-mass double detonation models from \cite{boos21a} as the initial conditions for our radiative transfer simulations; see Table \ref{tab:models} for model parameters.  The explosion models are calculated with the reactive hydrodynamics code \flash \citep{fryx00,dube09a}.  Most models are evolved until $\unit[100]{s}$, at which time they are in homologous expansion.  The initial core compositions are solar metallicity with a C/O mass ratio of $40/58.7$ and a $^{22}$Ne mass fraction of $0.013$.  The shells have initial mass fractions of 0.891 ($^4$He), 0.05 ($^{12}$C), 0.009 ($^{14}$N), and 0.05 ($^{16}$O), except for one $1.02 \msol$ model, which has a lower $^{16}$O mass fraction of 0.015 and a commensurately higher $^{4}$He mass fraction.  Nuclear burning is turned off in shocks and is artificially broadened in detonations with a burning limiter \citep{kush13a,kush20a,shen18a,boos21a}.  A 55-isotope nuclear reaction network is employed during the \flash simulations, and a 205-isotope network is used to post-process tracer particles; both networks are implemented using \texttt{MESA} \citep{paxt11,paxt13,paxt15a,paxt18a,paxt19a}.

\begin{figure}
  \centering
  \includegraphics[width=\columnwidth]{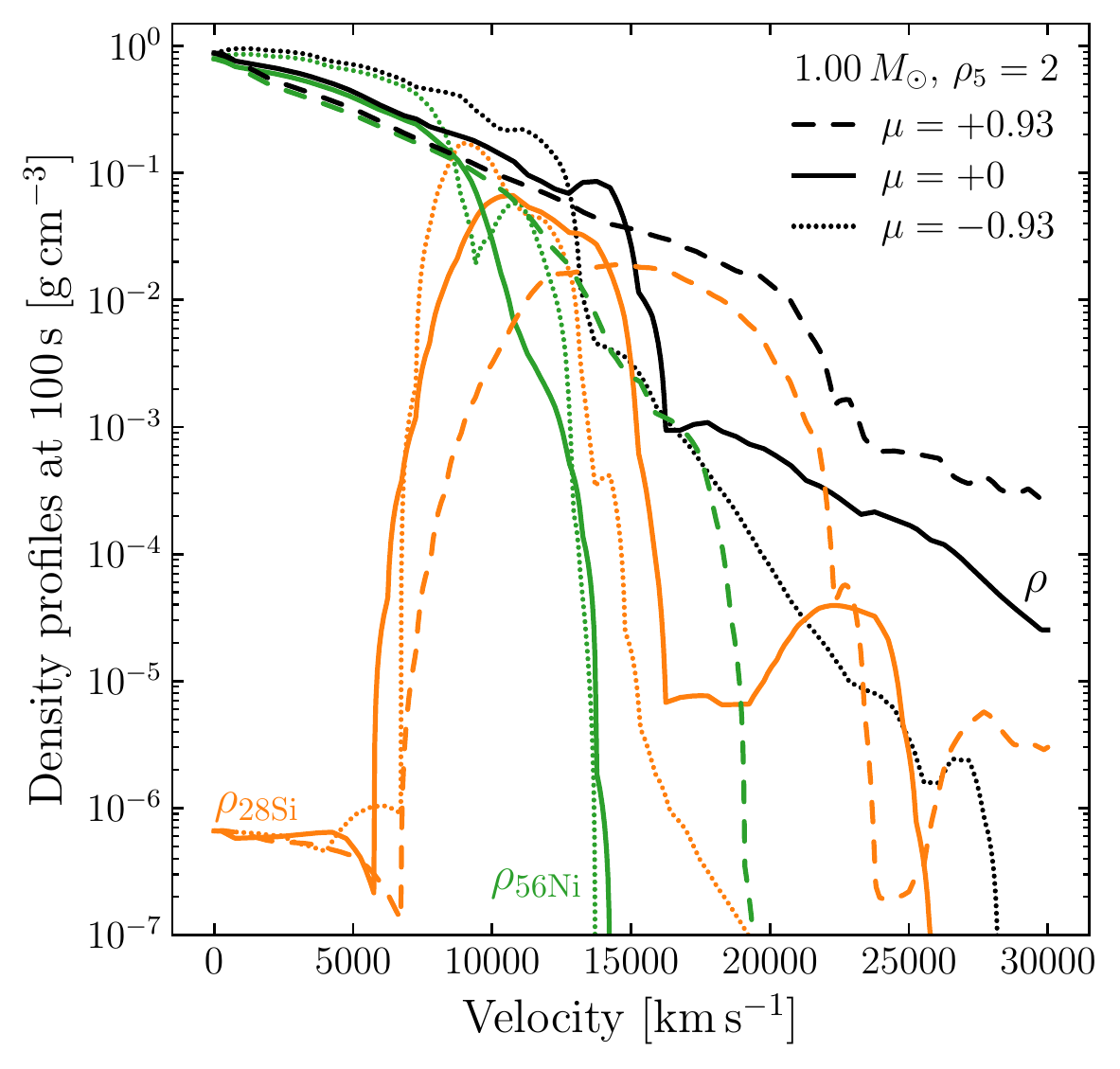}
  \caption{Profiles of the total mass density (black), $^{28}$Si mass density (orange), and $^{56}$Ni mass density (green) at $\unit[100]{s}$ for our $1.00 \msol$, $\rho_5=2$ explosion model.  Different line styles correspond to rays with different values of $\mu = +0.93$ (dashed), $0$ (solid), and $-0.93$ (dotted).}
  \label{fig:rays}
\end{figure}

Each model explosion is initiated in \flash via a hotspot along the symmetry axis in the helium shell.  We designate the hemisphere containing the hotspot as the northern hemisphere.  The ensuing helium detonation propagates around the surface of the WD towards the south pole, shedding a shock wave that travels into the core and eventually triggers a carbon core detonation near the symmetry axis in the southern hemisphere.  This second detonation propagates back out and burns the majority of the core.

Figure \ref{fig:rays} shows the resulting profiles of the total mass density, the $^{28}$Si mass density, and the $^{56}$Ni mass density for our $1.0 \msol$ model with an initial helium shell base density, normalized to $\unit[10^5]{g \, cm^{-3}}$, of $\rho_5 = \rho_{\rm base}/{\rm 10^5 \, g \, cm^{-3}}=2$.  Profiles are shown for three different values of $\mu = \cos \theta$, measured with respect to the symmetry axis in the northern hemisphere ($\mu = +1$ along the northern axis).  It is evident that there is significant asymmetry imparted by the multi-dimensional nature of the helium detonation and the off-center carbon core detonation.  In particular, as the carbon detonation propagates northwards from its birthplace in the southern hemisphere, it strengthens after it passes the center due to the negative density gradient and to the reduced effects of curvature.  As a result, $^{28}$Si is produced out to substantially higher velocities in the northern hemisphere, which will have a significant effect on the velocities inferred from spectra, which we discuss in Section \ref{sec:spectra}.  We refer the reader to \cite{boos21a} for further details on the explosion models used in this work.

The post-processed abundances, densities, and temperatures from the \flash simulations are interpolated onto two-dimensional axisymmetric velocity grids with $\unit[500]{km \, s^{-1}}$ resolution, which are then used as inputs for two-dimensional time-dependent radiation transport simulations with \sedona \citep{ktn06}.  \sedona is a Monte-Carlo-based radiative transfer code that has been used extensively to study a variety of astrophysical transients \citep{kase09a,kase09b,kase11a,shen10,wk11,barn13a,roth16a}.  The linelist is the same as that used in \cite{shen21a} and contains up to the first 1000 lowest energy levels for all ionization states of interest.  We treat line opacities in the ``expansion opacity'' formalism \citep{karp77a,east93a}.  We assume the absorption probability, $\epsilon$, is equal to a constant value of $1.0$ for all lines, which sets the line source functions equal to the Planck function.  This implies that all radiative excitations end up contributing to the thermal pool of photons.  We also assume that the ionization fractions and level populations are given by their local thermodynamic equilibrium (LTE) values.  The frequency grid is spaced evenly in logarithmic space, with a constant multiplicative factor of 1.0003 between gridpoints.  See \cite{shen21a} for further details of the effects of these choices on one-dimensional WD detonation models.


\section{Results}
\label{sec:results}

In this section, we present the light curves and spectra of our two-dimensional radiative transfer calculations and compare them  to observations.  Correlations among various quantities near maximum light are discussed in Section \ref{sec:maxcorr}.  Light curves and spectra are calculated along 15 viewing angles spaced evenly in $\mu = \cos \theta$, where $\theta$ is measured with respect to the symmetry axis.  The helium shell detonation is ignited in the northern hemisphere, where  $\mu > 0$,  and $\mu < 0$ in the southern hemisphere, where the carbon core detonation is born.  In this work, we present comparisons to observables from 10 days before to 3 days  after $B$-band maximum, due to computational constraints and our LTE assumption, which becomes less accurate at later times.  We will focus on modeling outside of this time frame in future studies.


\subsection{Light curves}

\begin{figure}
  \centering
  \includegraphics[width=\columnwidth]{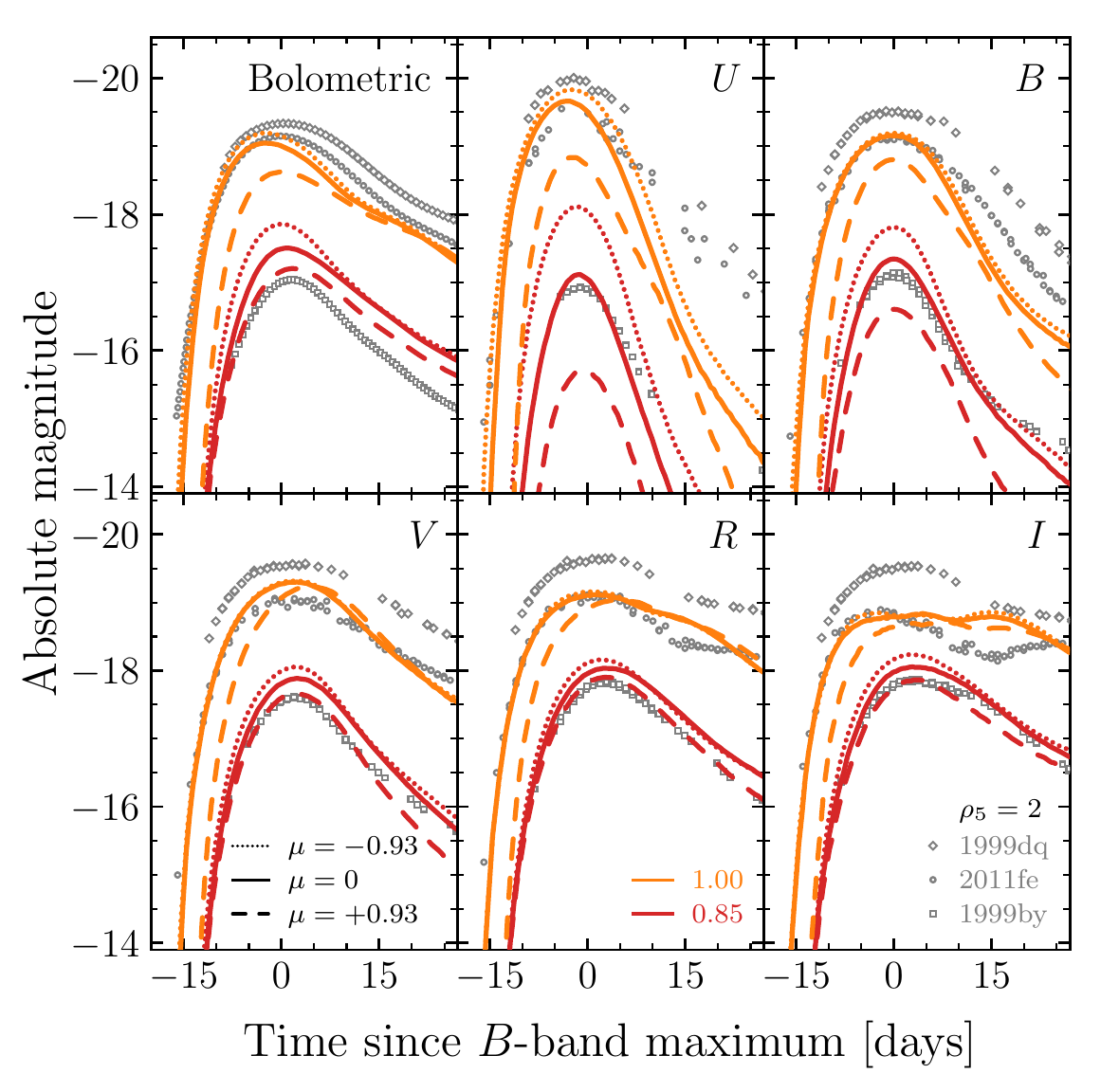}
  \caption{Multi-band light curves of our $0.85$ and $1.00 \msol$ models with  thin helium shells ($\rho_5=2$).  Three lines of sight ($\mu = -0.93$, $0$, and $+0.93$) are shown for each model, corresponding to different linestyles.  Gray points are observed SNe, as labeled; note that the observed ``bolometric'' light curves are more precisely quasi-bolometric light curves and are thus an underestimate of the true bolometric luminosities.}
  \label{fig:lcs_2e5_085_100}
\end{figure}

\begin{figure}
  \centering
  \includegraphics[width=\columnwidth]{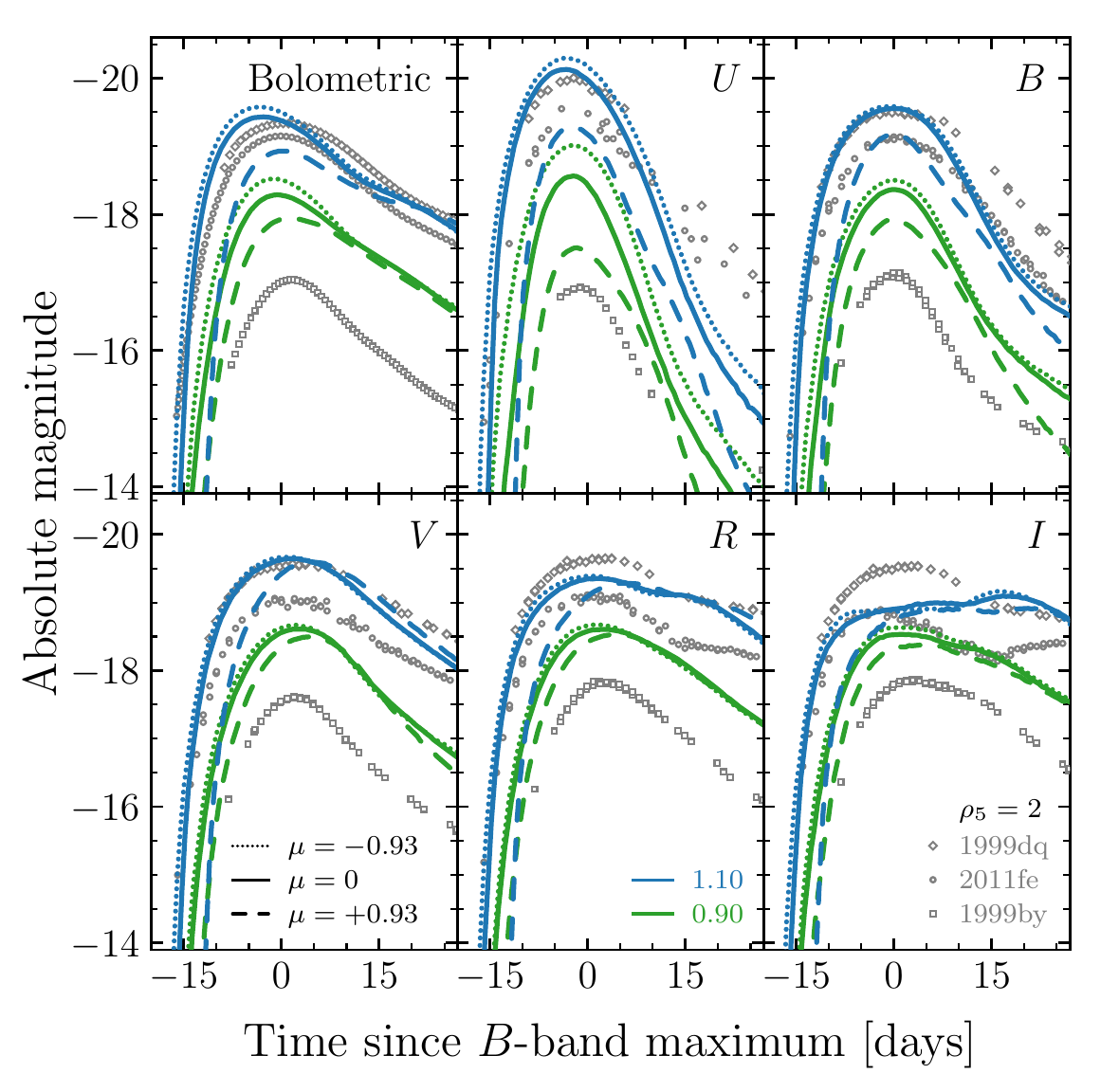}
  \caption{Same as Figure \ref{fig:lcs_2e5_085_100}, but for models with masses of $0.9$ and $1.1 \msol$.}
  \label{fig:lcs_2e5_090_110}
\end{figure}

Figures \ref{fig:lcs_2e5_085_100} and  \ref{fig:lcs_2e5_090_110} show multi-band light curves for models with total masses of $0.85$, $0.90$, $1.00$, and $1.10 \msol$ with the same initial helium shell base density of $\rho_5=2$.  Different linestyles correspond to different viewing angles of $  \mu = -0.93$, $0.0$, and $+0.93$, i.e., in the southern hemisphere, along the equator, and in the northern hemisphere, respectively.  We do not show the light curves of the $1.02 \msol$, $\rho_5=2$ models for clarity.  They are similar to the light curves of the $1.00 \msol$ model, but are slightly more luminous (the $B$-band maxima from different lines of sight are at most $\unit[0.1]{mag}$ brighter) as befits their somewhat higher $^{56}$Ni production ($0.53 \msol$ vs.\ $0.50 \msol$).

Gray symbols in both figures represent photometry of subluminous SN~1999by  \citep{garn04,stri05a,gane10a}, normal SN~2011fe \citep{muna13b,pere13a,tsve13a}, and overluminous SN~1999dq \citep{stri05a,jha06b,gane10a}.\footnote{Some of the observational data used in this work was obtained through \texttt{https://sne.space} \citep{guil17a}.}   SN~1999dq's light curve has been corrected for Milky Way reddening \citep{schl11a}.  Note that the observed ``bolometric'' light curves  shown here and throughout this work are actually quasi-bolometric light curves, which neglect the flux bluer than the $U$-band and redder than the $I$-band.   For the normal SN~2011fe and overluminous SN~1999dq, the quasi-bolometric and true bolometric luminosities are not significantly different \citep{sunt96a}.  However, for the subluminous SN~1999by, this may be as much as a 50\% underestimate of the luminosity at later times, as we discuss below.  We do not attempt to produce quasi-bolometric light curves from our models for comparison because they are significantly affected by our LTE approximation.

We note that non-LTE effects also have a significant impact on light curves after maximum light \citep{shen21a}, especially in the $U$-, $B$-, and $I$-bands.  We show the evolution of the light curves well past maximum for completeness, but we limit our post-maximum quantitative analysis to the (true) bolometric light curves, which should be relatively immune from non-LTE corrections.

The equatorial light curves for the $0.85$, $1.0$, and $1.1 \msol$ models provide a reasonable fit  in all bands to the three sets of observed SN photometry up to maximum light, with models of increasing mass matching observations of increasingly luminous SNe.  The largest discrepancy is in the $I$-band comparison of the $1.1 \msol$ model and SN~1999dq.  We note that a similar difference is seen in the one-dimensional non-LTE models of \cite{shen21a}.

The light curves along varying lines of sight are markedly different, particularly in the bluer bands, and especially for the lowest mass model, for which the maximum $B$-band magnitudes vary by $> \unit[1]{mag}$.  The variation in maximum magnitudes is not as significant for the higher mass models, but it is large enough that the brightest $1.0 \msol$ model line of sight reaches the same magnitude as the dimmest $1.1 \msol$ model line of sight.  There is a similar disparity in the timescales to reach maximum light at different viewing angles.  The most rapidly rising viewing angles for the  $1.0$ and $1.1 \msol$ models reach maximum light in nearly the same time that the most slowly rising $0.85 \msol$ model line of sight does.

\begin{figure}
  \centering
  \includegraphics[width=\columnwidth]{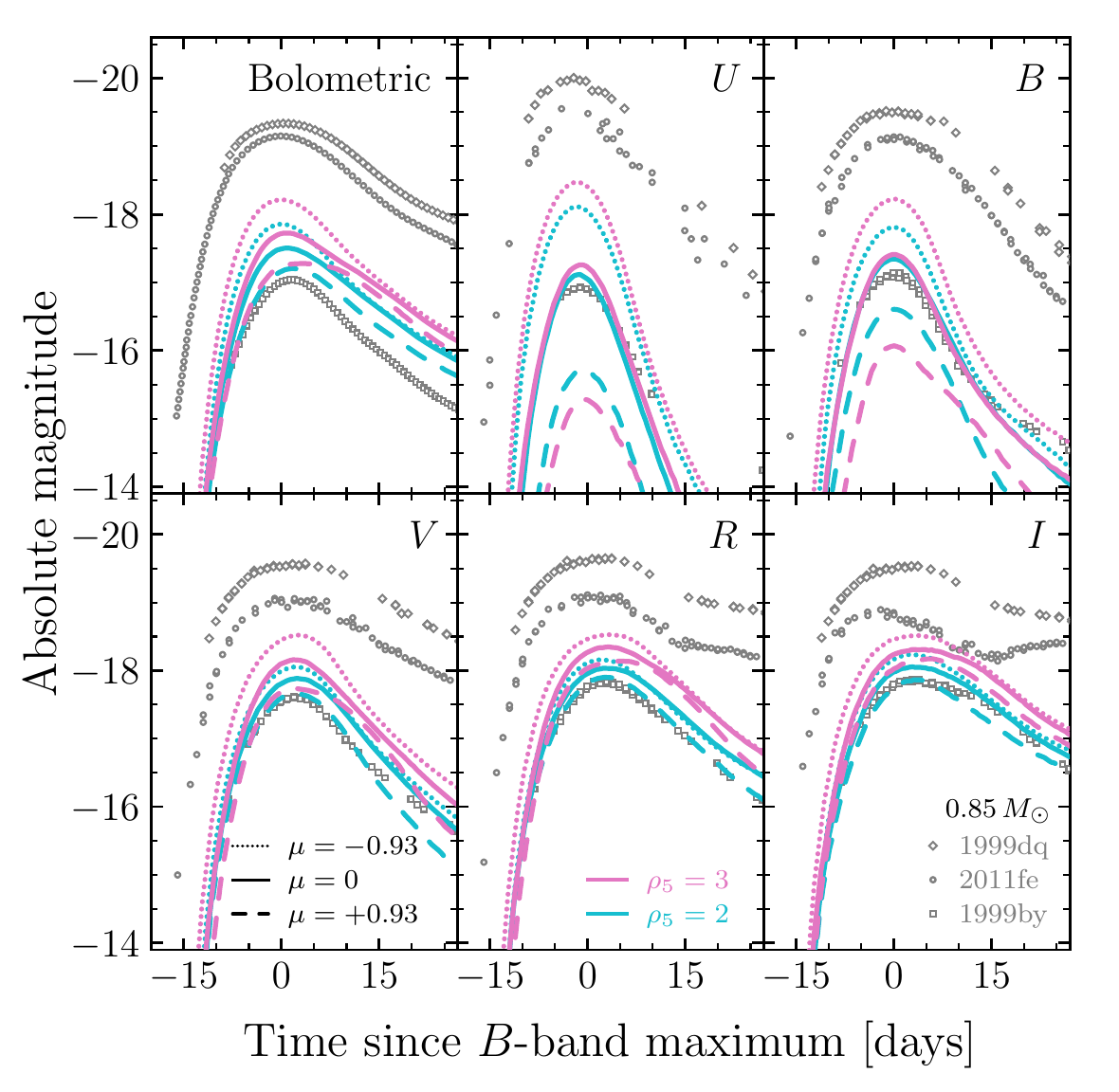}
  \caption{Same as Figure \ref{fig:lcs_2e5_085_100}, but for $0.85 \msol$ models with $\rho_5=2$ and $3$.}
  \label{fig:lcs_085}
\end{figure}

\begin{figure}
  \centering
  \includegraphics[width=\columnwidth]{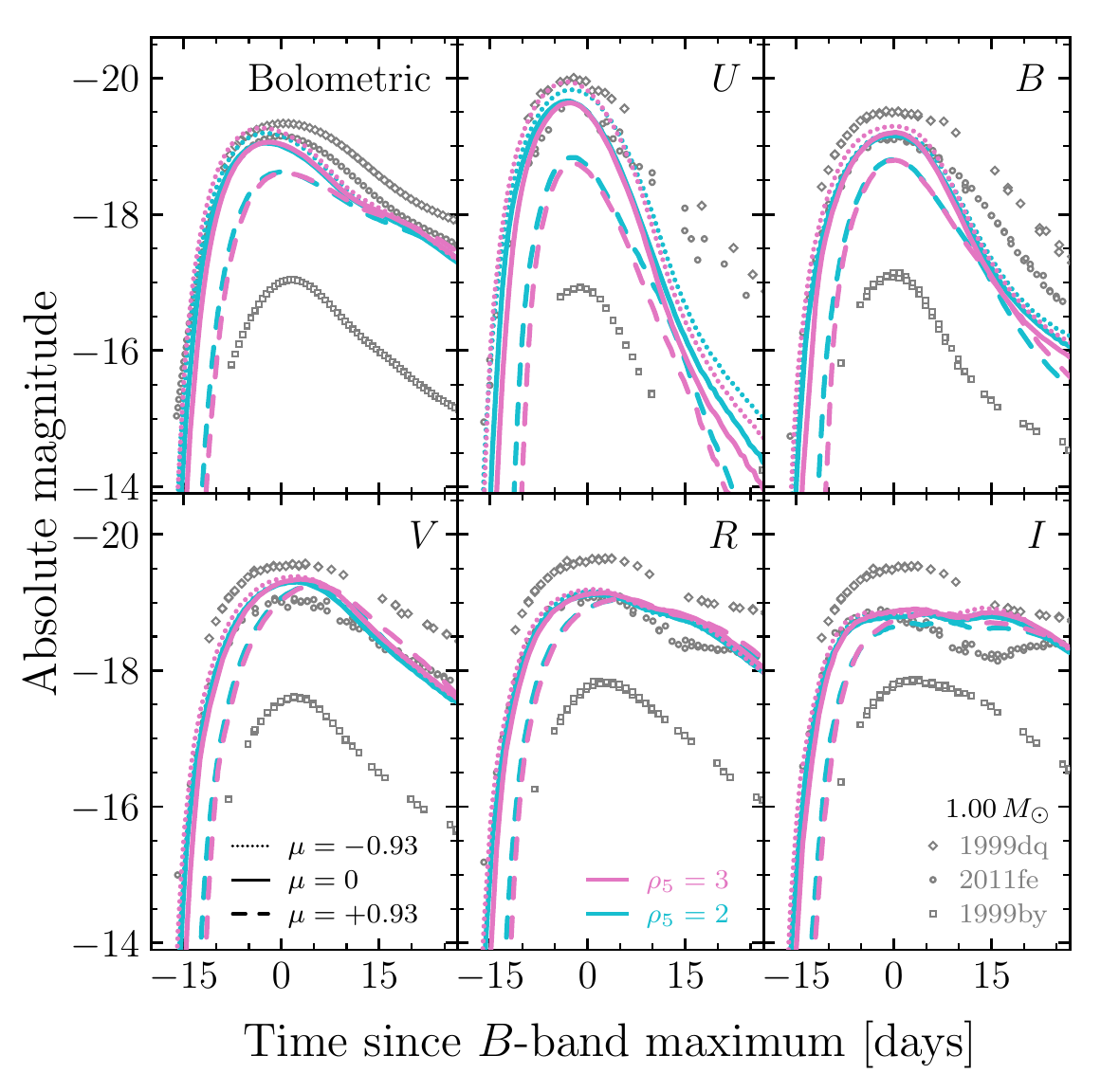}
  \caption{Same as Figure \ref{fig:lcs_085}, but for $1.00 \msol$ models.}
  \label{fig:lcs_100}
\end{figure}

\begin{figure}
  \centering
  \includegraphics[width=\columnwidth]{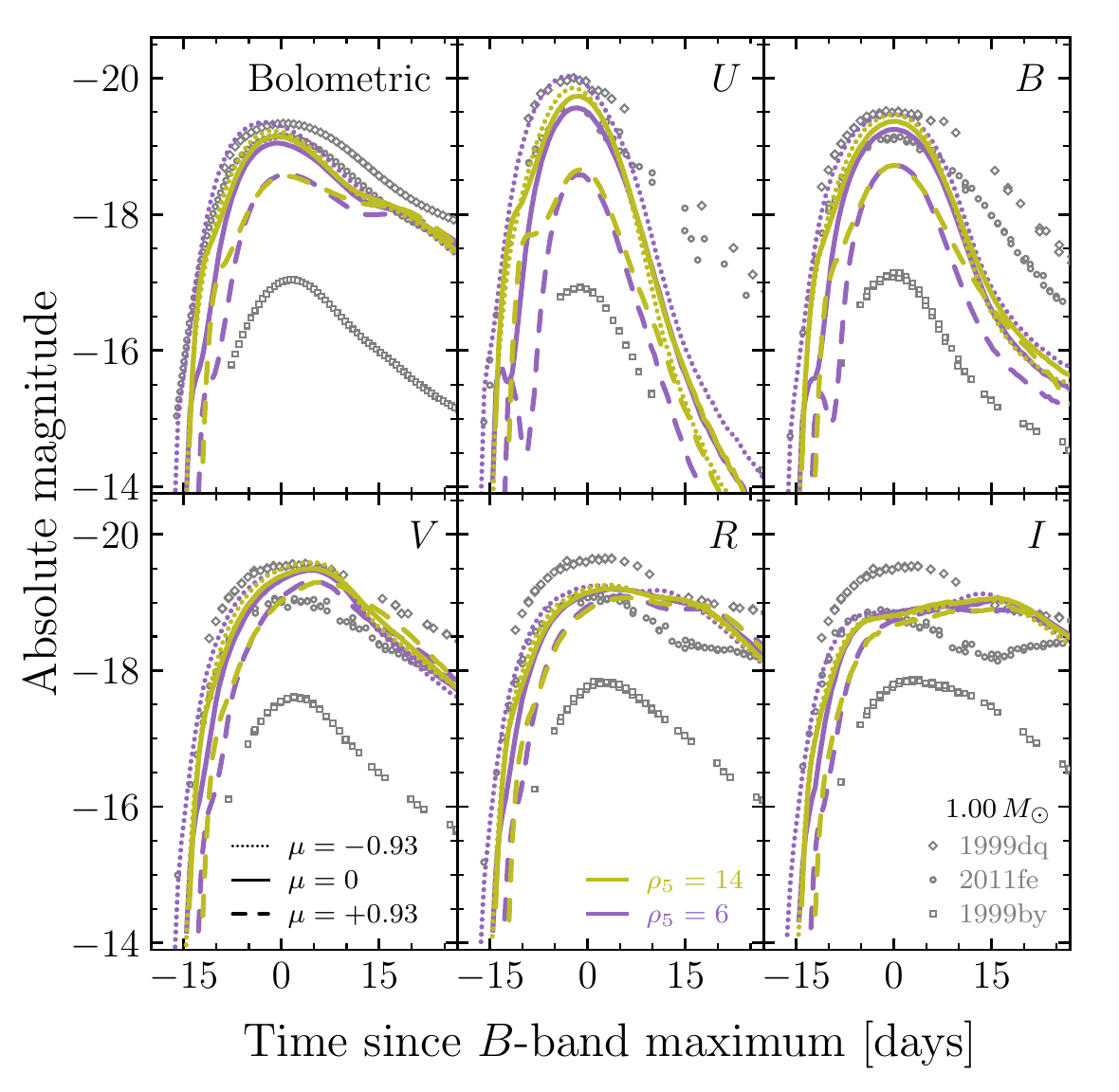}
  \caption{Same as Figure \ref{fig:lcs_100}, but for base densities of $\rho_5=6$ and $14$.}
  \label{fig:lcs_100_high}
\end{figure}

\begin{figure}
  \centering
  \includegraphics[width=\columnwidth]{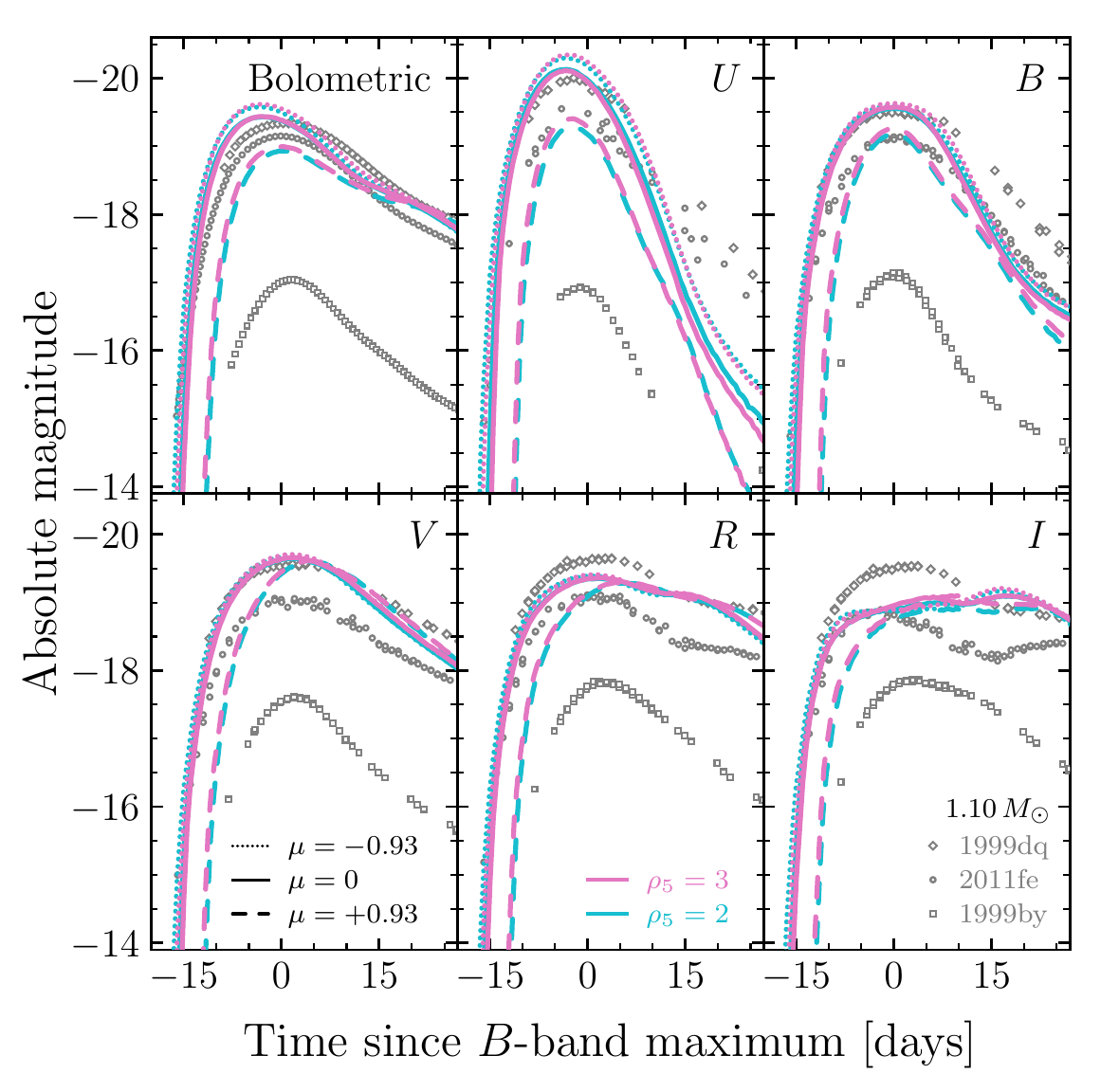}
  \caption{Same as Figure \ref{fig:lcs_085}, but for $1.10 \msol$ models.}
  \label{fig:lcs_110}
\end{figure}

We compare models with the same total masses but different helium shell base densities  in Figures \ref{fig:lcs_085} -- \ref{fig:lcs_110}.  For the $1.0$ and $1.1 \msol$ models with $\rho_5=2$ and $3$ (Figures \ref{fig:lcs_100} and \ref{fig:lcs_110}), the differences among the light curves are relatively minimal.  Peak magnitudes and light curve shapes are nearly identical for all viewing angles.  However, the differences are more significant for the $0.85 \msol$ models shown in Figure \ref{fig:lcs_085}: the $\rho_5=2$ model is brighter in $B$-band than the $\rho_5=3$ model when viewed from the northern hemisphere, where the helium shell detonation is ignited, possibly due to the much larger amount of Ti and associated line-blanketing in the northern shell of the thicker helium shell model, but is dimmer when viewed from the southern hemisphere, likely due to the larger amount of $^{56}$Ni produced by the $\rho_5=3$ model.

There are also significant differences for the thick helium shell $1.0 \msol$ models shown in Figure \ref{fig:lcs_100_high}.  While the amount of radioactive $^{48}$Cr and $^{56}$Ni produced in the helium shell is very low for the $1.0 \msol$, $\rho_5=2$ and $3$ models ($< 2\E{-5} \msol$ for both), the $\rho_5=6$ and $14$ models produce significantly more radioactive material: $7\E{-3}$ and $0.05 \msol$, respectively.  The radioactive decay of this material near the surface of the ejecta leads to early ``shoulders'' and ``humps'' in the light curves of these models.  Similar features are seen in previous studies of one-dimensional relatively massive helium shell models \citep{ww94,poli19a}, which have been invoked to explain several peculiar SNe~Ia \citep{de19a,jaco20a}.  The present paper focuses on thin-shell models, but we include results from these two thick-shell models for completeness; we leave a more thorough examination of models with massive helium shells to future work.

\begin{figure}
  \centering
  \includegraphics[width=\columnwidth]{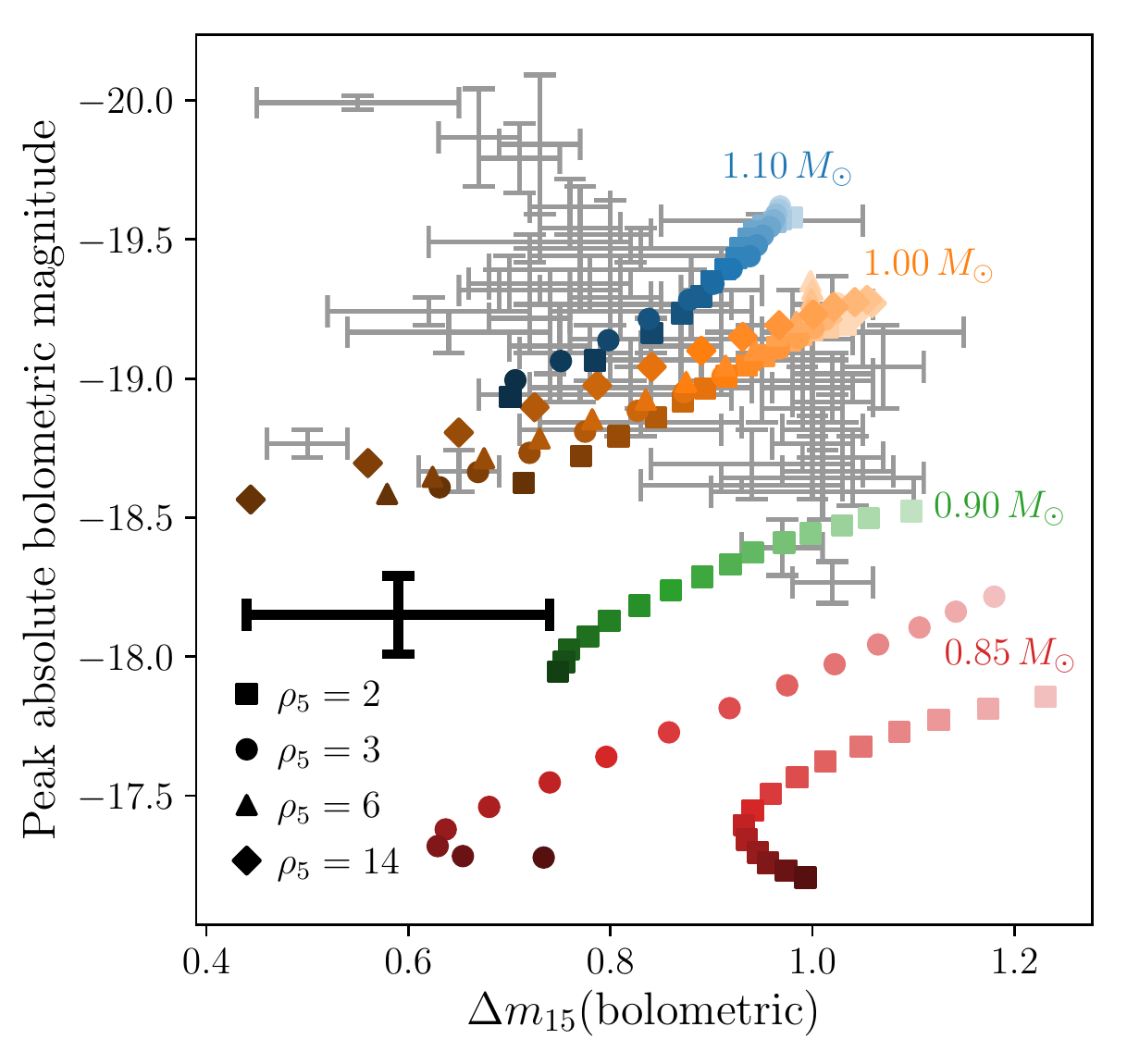}
  \caption{Peak absolute bolometric magnitude vs.\ $\Delta m_{15}({\rm bolometric})$, i.e., the bolometric version of the \cite{phil93a} relation.  Gray error bars are observational data from \cite{scal19a}, and results from our theoretical models are as labeled.  Light to dark colors are different lines of sight with increasing values of  $  \mu$ from $-0.93$ to $+0.93$, equally spaced in $ \mu $.  The black error bar shows the maximum differences between the LTE and non-LTE results from \cite{shen21a} for the relevant quantities.}
  \label{fig:phillipsbol}
\end{figure}

Figure \ref{fig:phillipsbol} shows the bolometric version of the \cite{phil93a} relationship, comparing the maximum bolometric magnitude to the 15-day decline in bolometric magnitude following the time of maximum,  $\Delta m_{15}({\rm bolometric})$.  (Given the differences between LTE and non-LTE results presented in \citealt{shen21a}, we do not present analogous comparisons in any broad-band filters.)  Theoretical models are as labeled, with lines of sight from southern to northern hemispheres corresponding to symbols of increasing darkness.  Observed values from \cite{scal19a} are in gray.  The black error bar shows the differences in bolometric peak magnitude and in $\Delta m_{15}({\rm bolometric})$ between LTE and non-LTE radiative transfer calculations of one-dimensional sub-Chandraskehar-mass WD detonations from \cite{shen21a}.  These differences should not be thought of as directly applicable to the models we discuss in this paper, since the one-dimensional bare WD simulations analyzed in \cite{shen21a} are obviously different from multi-dimensional thin-shell double detonation models, but they may be suggestive of the possible changes when non-LTE multi-dimensional radiation transport is performed in the future.

Most of the lines of sight for the $1.0$ and $1.1 \msol$ models yield results within the observed range.  However, the northernmost lines of sight for the $1.0 \msol$ models with higher helium shell base densities ($\rho_5 \geq 3$) lead to light curves that fade too slowly for their peak luminosity.  There are two SNe~Ia with similar parameters, but they are the peculiar SN~2006bt and SN~2006ot \citep{fole10b,stri11a}.  Even given the possible variance between LTE and non-LTE models suggested by the black error bar, double detonations with $\rho_5 \gtrsim 6$ helium shells appear to be ruled out as the dominant progenitors of SNe~Ia.

The \cite{scal19a} observed sample contains very few SNe~Ia with bolometric luminosities as low as those produced by our $0.9 \msol$ model due to the magnitude-limited nature of the surveys from which they draw.  The few observed low-luminosity SNe that do exist in the sample are consistent with the bolometric decline rates of the southern viewing angles.  However, similar to the $1.0$ and $1.1 \msol$ thick shell models, the $0.9 \msol$ model's northernmost lines of sight do not appear to match SNe found in nature, with the repeated caveat that the observed sample size is small.

None of the observed light curves in the sample probes the faintest end of the SN~Ia luminosity distribution, with similar bolometric magnitudes to those of the low-mass $0.85 \msol$ models.\footnote{The sample does include SN~2006gt and SN~2007ba, which are classified spectroscopically as SN~1991bg-likes, but they both have absolute peak $B$-band magnitudes of  $-18.1$ even before correcting for extinction, which is significantly higher than the peak $B$-band magnitudes of truly faint SNe like SN~1991bg ($-16.9$) and SN~1999by ($-17.2$).}  We have examined some of the SN~1991bg-like SNe in the literature but did not find any truly bolometric results.  Several studies do report quasi-bolometric results (e.g., \citealt{stri05a} and \citealt{taub08a}), integrating the flux in the $U$-, $B$-,  $V$-,  $R$-,  and $I$-bands; however, the quasi-bolometric to bolometric flux ratios of our low-mass models evolve significantly with time and become as low as 50\% 15 days after the time of bolometric maximum.  Furthermore, while the true bolometric light curve is relatively immune from non-LTE corrections, the quasi-bolometric light curve is much more sensitive to non-LTE effects, due to the redistribution of flux into the Ca {\sc ii} near-infrared triplet feature at wavelengths where the transmission of the $I$-band begins to decrease.  Thus, we do not attempt to match our low-mass models to the quasi-bolometric data in the literature.


\subsection{Spectra}
\label{sec:spectra}

\begin{figure}
  \centering
  \includegraphics[width=\columnwidth]{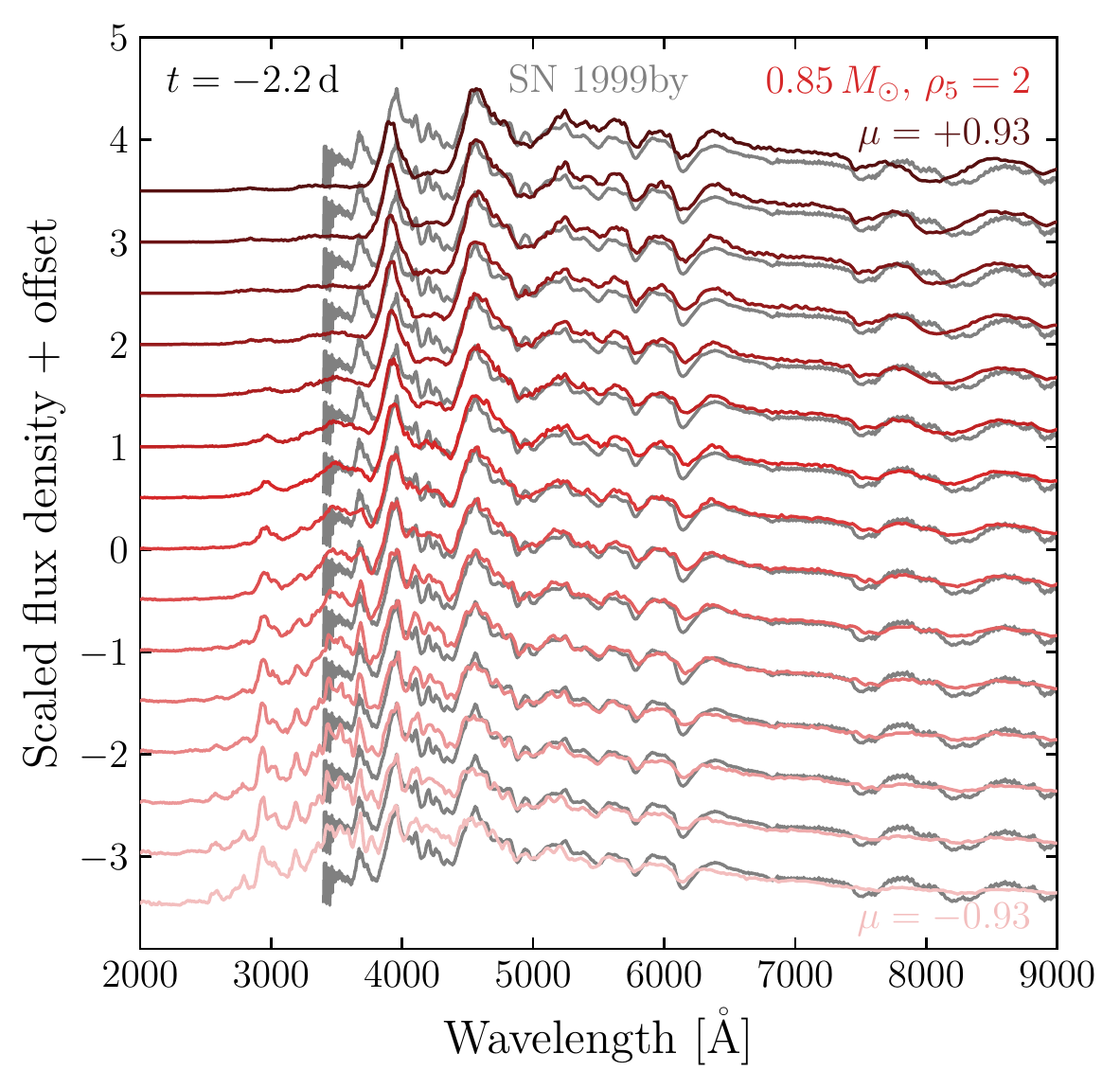}
  \caption{Spectra of the $0.85 \msol$, $\rho_5=2$ model from different lines of sight at $-2.2$ days from the time of $B$-band maximum.  Values of $ \mu$ decrease from top to bottom.  Gray lines are the spectrum of SN~1999by at $-2.2$ days, scaled by an arbitrary constant and reproduced 15 times for ease of comparison.}
  \label{fig:spec_mu_085_2}
\end{figure}

\begin{figure}
  \centering
  \includegraphics[width=\columnwidth]{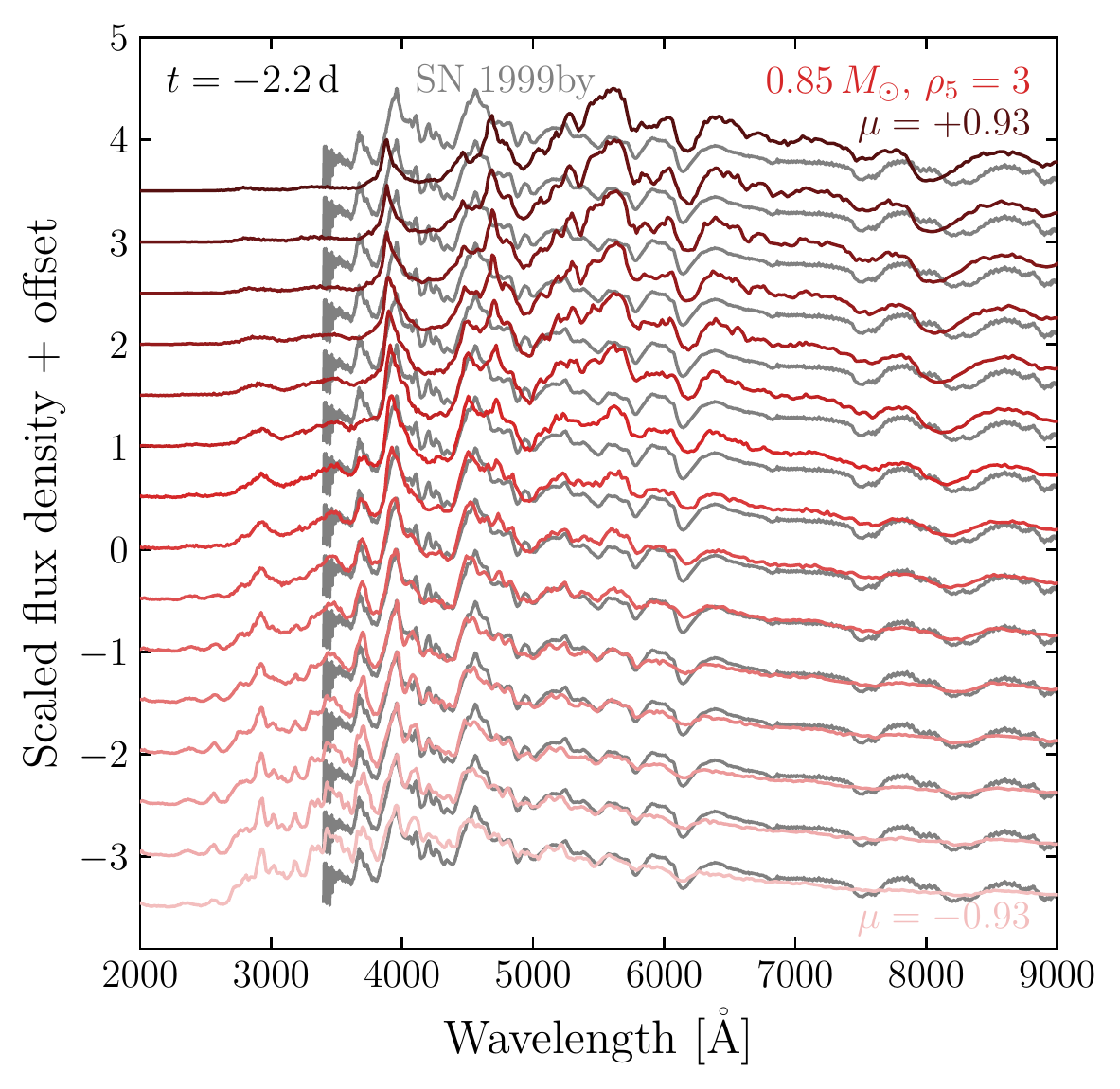}
  \caption{Same as Figure \ref{fig:spec_mu_085_2}, but for the model with $\rho_5=3$.}
  \label{fig:spec_mu_085_3}
\end{figure}

\begin{figure}
  \centering
  \includegraphics[width=\columnwidth]{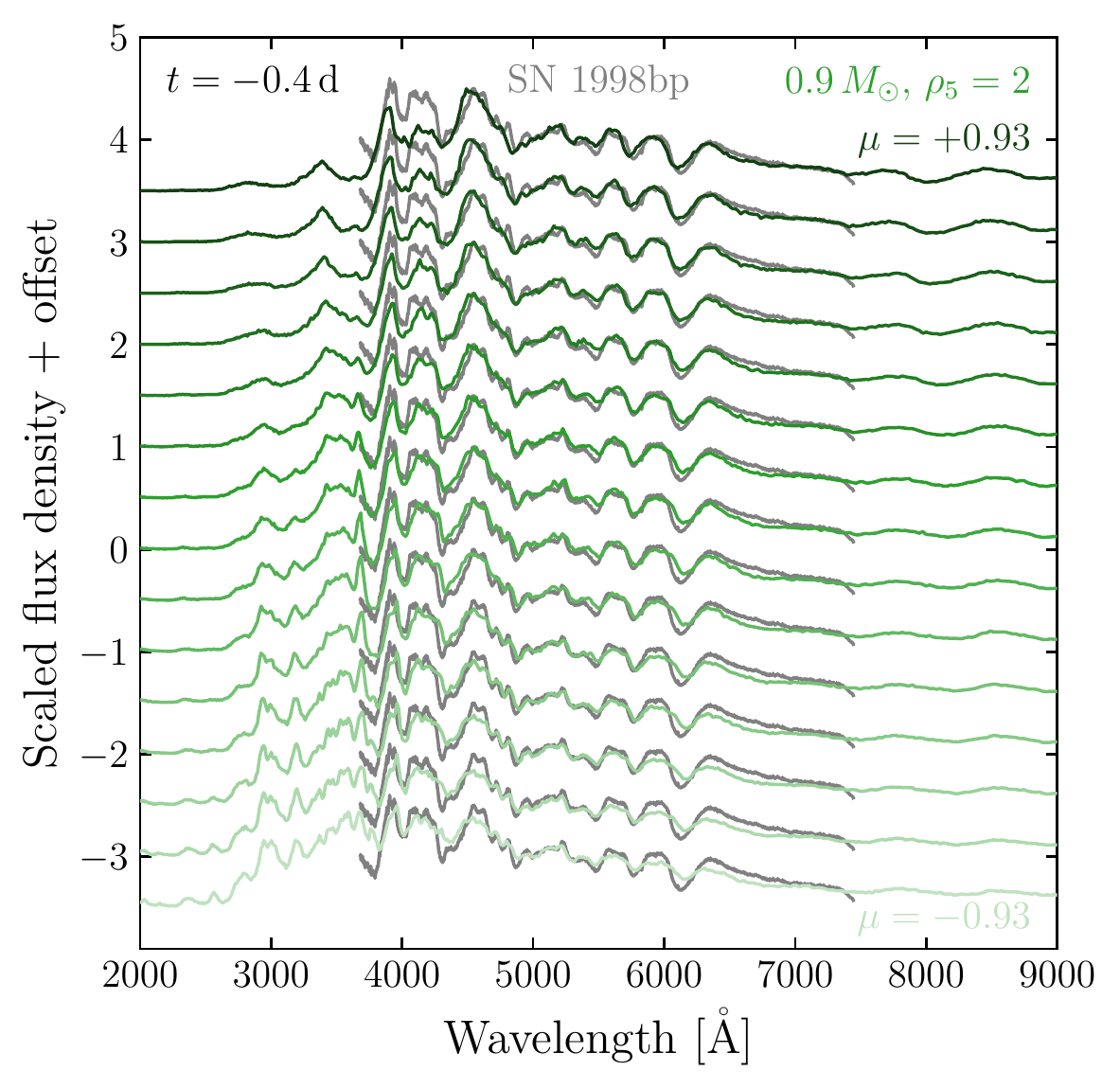}
  \caption{Same as Figure \ref{fig:spec_mu_085_2}, but for the $0.90 \msol$, $\rho_5=2$ model $-0.4$ days from the time of $B$-band maximum.  Gray lines are the spectrum of SN~1998bp at the same time.}
  \label{fig:spec_mu_090_2}
\end{figure}

\begin{figure}
  \centering
  \includegraphics[width=\columnwidth]{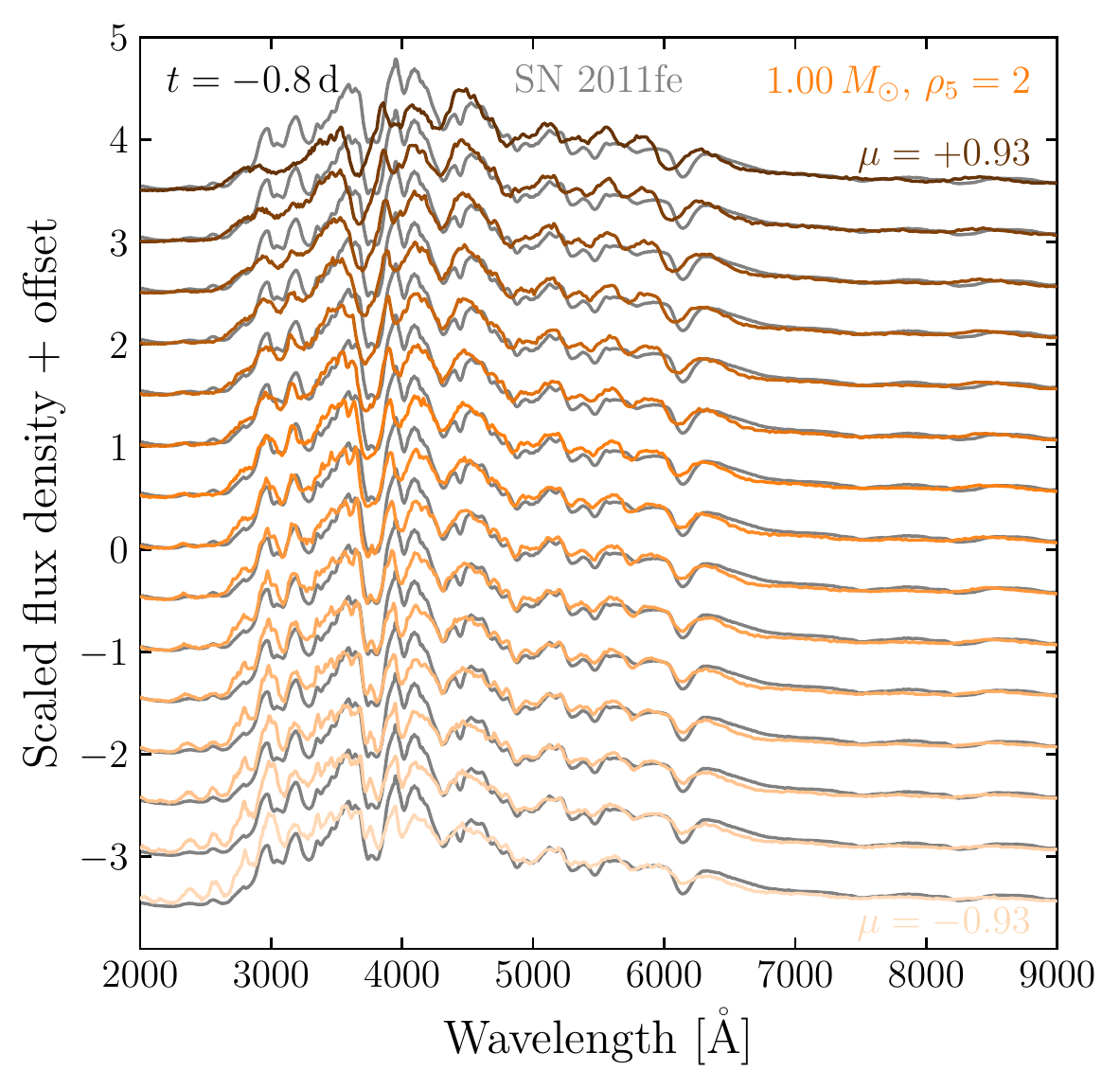}
  \caption{Same as Figure \ref{fig:spec_mu_085_2}, but for the $1.00 \msol$, $\rho_5=2$ model $-0.8$ days from the time of $B$-band maximum.  Gray lines are the spectrum of SN~2011fe at the same time.}
  \label{fig:spec_mu_100_2}
\end{figure}

\begin{figure}
  \centering
  \includegraphics[width=\columnwidth]{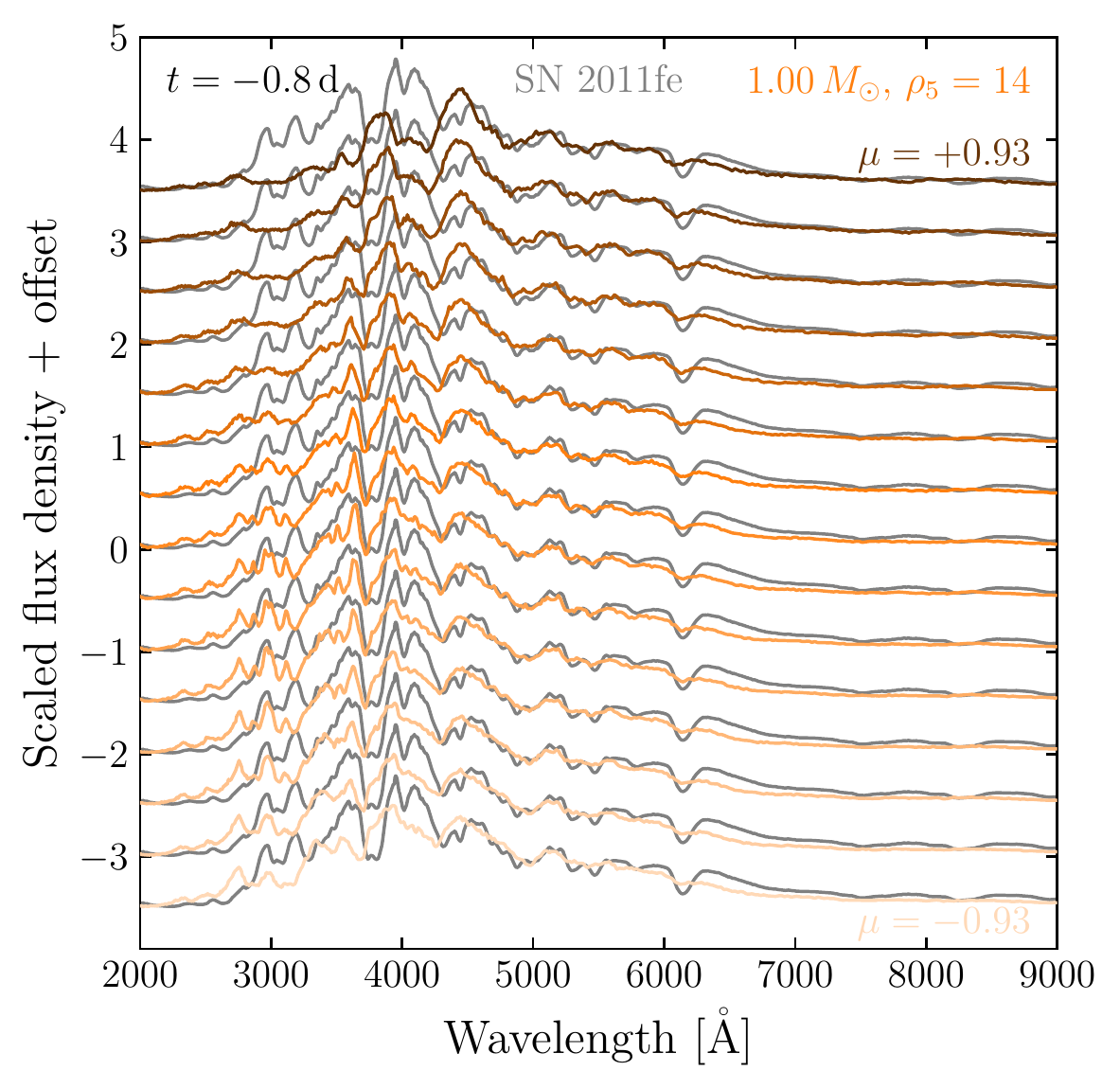}
  \caption{Same as Figure \ref{fig:spec_mu_100_2}, but for the model with $\rho_5=14$.}
  \label{fig:spec_mu_100_14}
\end{figure}

\begin{figure}
  \centering
  \includegraphics[width=\columnwidth]{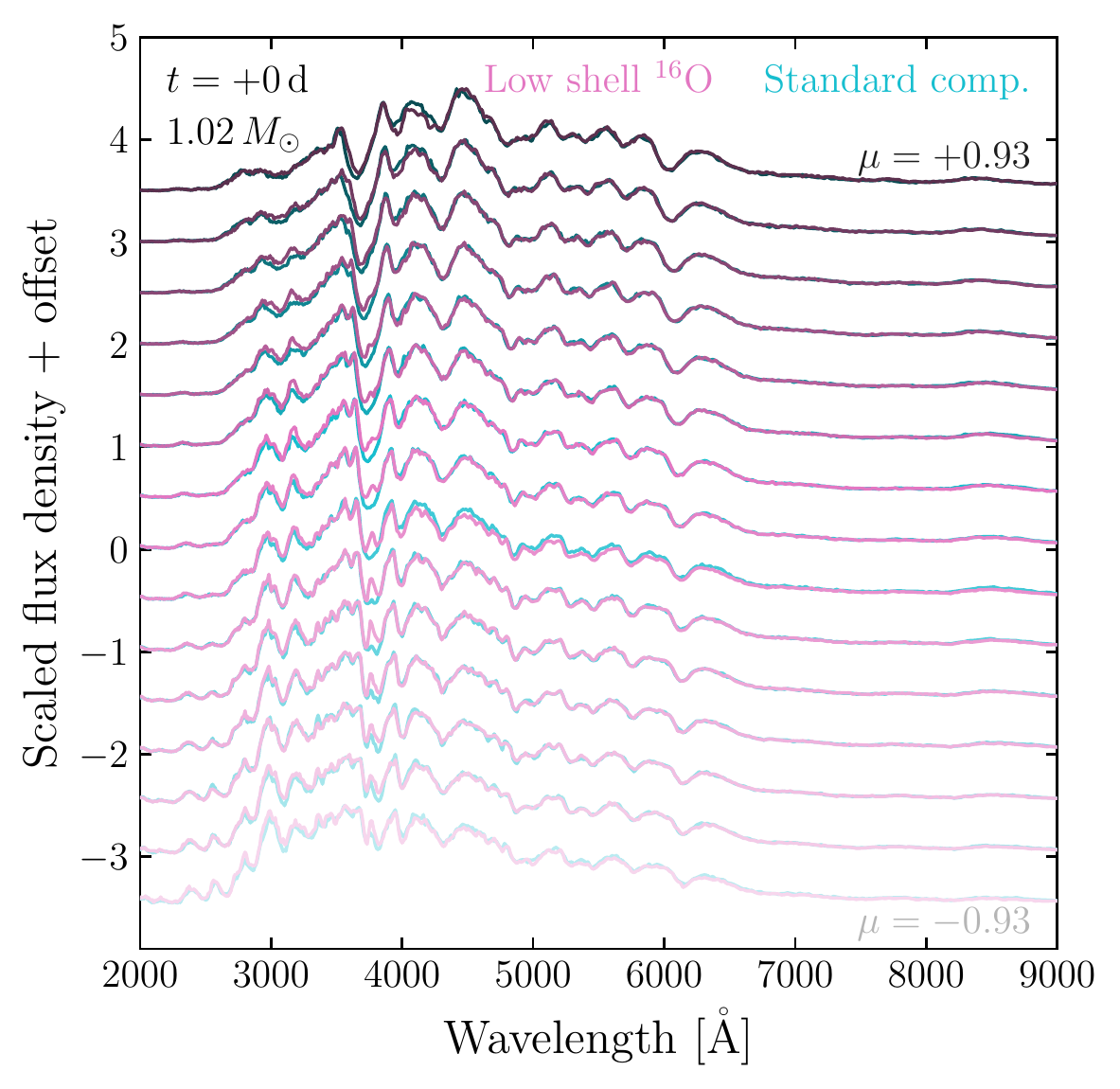}
  \caption{Same as Figure \ref{fig:spec_mu_085_2}, but for the $1.02 \msol$, $\rho_5=2$ models $+0.0$ days from the time of $B$-band maximum.  Cyan lines show spectra of the model with a standard initial helium shell composition; magenta lines show spectra for the model with a reduced $^{16}$O initial shell abundance.}
  \label{fig:spec_mu_102_2}
\end{figure}

\begin{figure}
  \centering
  \includegraphics[width=\columnwidth]{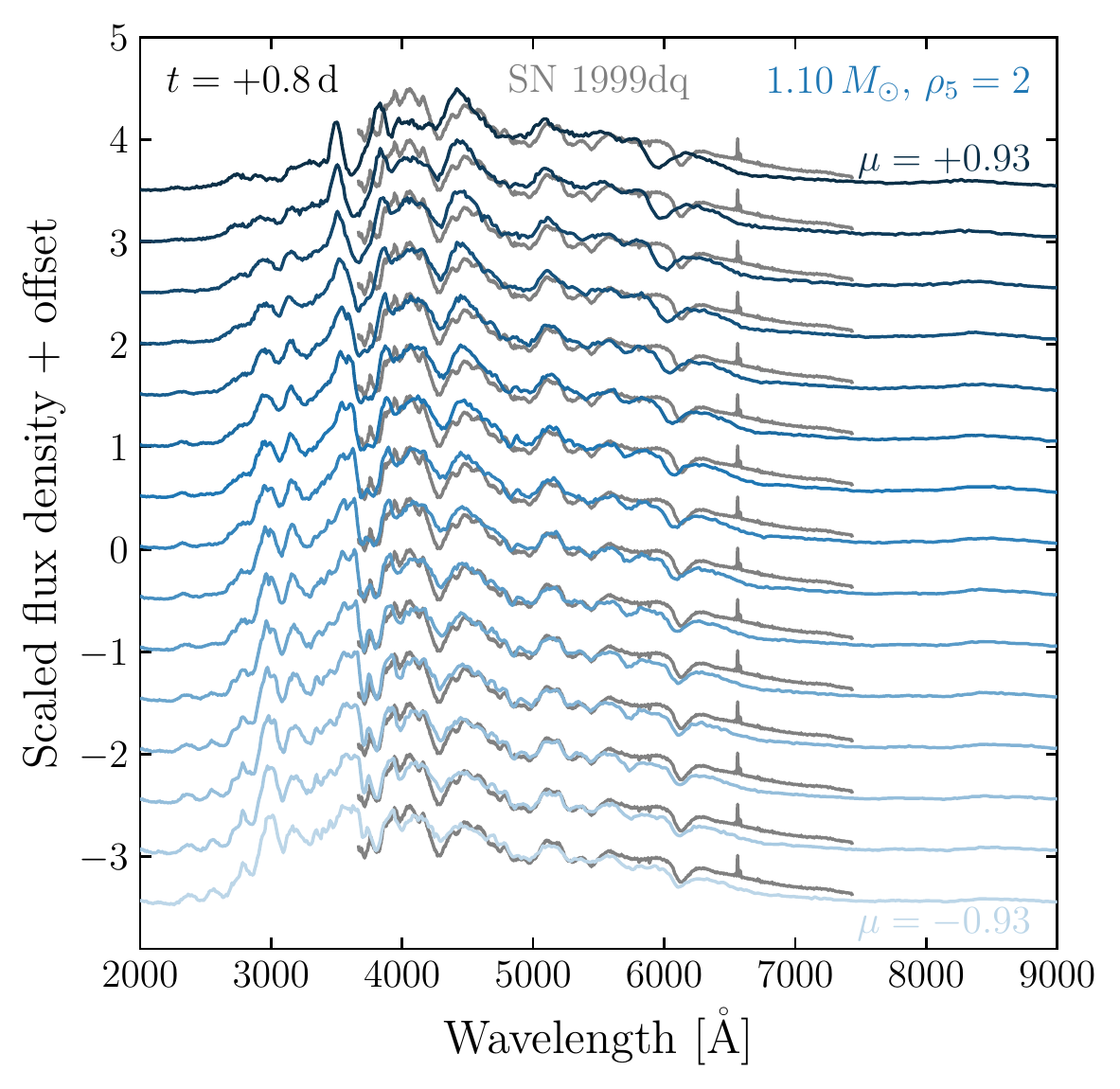}
  \caption{Same as Figure \ref{fig:spec_mu_085_2}, but for the $1.10 \msol$, $\rho_5=2$ model $+0.8$ days from the time of $B$-band maximum.  Gray lines are the spectrum of SN~1999dq at the same time.}
  \label{fig:spec_mu_110_2}
\end{figure}

In Figures \ref{fig:spec_mu_085_2} -- \ref{fig:spec_mu_110_2}, we show near-maximum-light spectra of a subset of our models from different viewing angles, scaled so that the difference between the maximum and minimum of each spectrum is equal to 1.  We also show spectra of observed SNe at the same phase with respect to the time of $B$-band maximum, chosen to approximately match the light curve evolution viewed along the equatorial line of sight of each model.  The observed spectra are scaled by an arbitrary factor that is constant within each figure but differs between figures.

As Figure \ref{fig:spec_mu_085_2} shows, the subluminous SN~1999by \citep{math08a} provides a particularly good fit to the equatorial spectrum of the $0.85 \msol$, $\rho_5=2$ model and satisfactorily matches the spectra further into the southern hemisphere (the hemisphere where the carbon core detonation is ignited).  However, in the northern hemisphere, the Ti {\sc ii} trough located near $\unit[4000]{\AA}$ becomes very deep, due to the lower temperatures caused by both the more radially extended distribution of Ti and the lower flux along these lines of sight.  Furthermore, the observed wavelength of the Si {\sc ii} $\lambda 6355$ absorption minimum is slightly blueshifted with respect to the models; we discuss this further in Section \ref{sec:maxcorr}.  The overall fits to the $0.85 \msol$, $\rho_5=3$ model lines of sight shown in Figure \ref{fig:spec_mu_085_3} are not as good as those for the $\rho_5=2$ model, matching expectations from the larger color deviations seen in Figure \ref{fig:lcs_085}, and a similar mismatch of the Ti {\sc ii} trough is seen in the northern hemisphere of this model as well.

Our $0.9 \msol$ model is compared to the transitional SN~1998bp \citep{math08a}, corrected for Milky Way reddening \citep{schl11a}, in Figure \ref{fig:spec_mu_090_2}.  The fits to the lines of sight near the equator are particularly good.  In the southern hemisphere (fainter lines), the model absorption features throughout the optical are not quite as deep as for the observed spectrum; in the northern hemisphere (darker lines), the model absorption complex near $\unit[4000]{\AA}$, containing Mg {\sc ii}, Si {\sc ii}, Fe {\sc ii},  and possibly Ti {\sc ii} lines, is too deep.  

For this and the higher-mass $1.0$ and $1.1 \msol$ models, there is a strong Si {\sc ii} velocity dependence on the viewing angle: velocities in the southern hemisphere are roughly similar to each other, and they increase as the viewing angle moves northwards.  This trend matches the qualitative features seen in Figure \ref{fig:rays}, in which the southern and equatorial $^{28}$Si density profiles are relatively similar, while the northern density profile extends to higher velocities.  A similar distribution of velocities with respect to viewing angle was found in \cite{town19a}'s radiative transfer calculations  and was used by \cite{zhan20a} to provide a broad match to the observed SN~Ia velocity distribution.

Normal SN~2011fe \citep{mazz14a} is compared to the lowest and highest helium shell base density $1.0 \msol$ models in Figures \ref{fig:spec_mu_100_2} and \ref{fig:spec_mu_100_14}.  As before, the equatorial line of sight for the lowest helium shell density model yields a reasonable fit to the observed spectrum, although there are some discrepancies (e.g., the depth of the Si {\sc ii} $\lambda 4130$ feature and the locations of the Si {\sc ii} $\lambda 4130$ and $\lambda 6355$ absorption minima).  There is similar agreement throughout the southern hemisphere.  However, the spectra viewed from the north deviate significantly from that of SN~2011fe, particularly when comparing the velocities of the features: the model velocities are blueshifted by up to $\unit[4000]{km \, s^{-1}}$ for the northernmost line of sight.

Figure \ref{fig:spec_mu_100_14} compares SN~2011fe's near-maximum-light spectrum to those of the $1.0 \msol$, thickest helium shell ($\rho_5=14$) model.  While there is some agreement at the equator and in the southern hemisphere, there are larger discrepancies than for the thin shell ($\rho_5=2$) model.  In the northern hemisphere, the velocity disagreements are not as striking as for the thin shell case, but the model spectra have much more absorption below $\unit[4000]{\AA}$, commensurate with the mismatch in $U-B$ colors seen in Figure \ref{fig:lcs_100_high}.

We examine the effects of changing the initial $^{16}$O shell abundance in Figure \ref{fig:spec_mu_102_2}.  Cyan lines show spectra for a $1.02 \msol$, $\rho_5=2$ model with an initial helium shell composition identical to that of the models with different masses.  Magenta lines show spectra for a $1.02 \msol$ model with a reduced $^{16}$O mass fraction in the shell of $0.015$.  The spectra are nearly identical from all viewing angles, except for a slight deviation in the Ca {\sc ii} H\&K feature near the equator: the model with a reduced $^{16}$O shell abundance does not have as strong an absorption feature as the standard abundance model, due in part to a two-fold reduction in the amount of Ca that is synthesized in the shell.

In Figure \ref{fig:spec_mu_110_2}, we compare the near-maximum-light spectra of our $1.1 \msol$, $\rho_5=2$ model to that of the overluminous SN~1999dq \citep{math08a}, which is corrected for Milky Way reddening \citep{schl11a}.  The equatorial line of sight is somewhat less of a satisfactory fit than for the previous comparisons, but the model still captures the basic features of the observed spectrum.  However, there is a significant discrepancy in the Si {\sc ii} $\lambda 6355$ velocity, which is even more striking in the northern hemisphere; the observed spectrum's Si {\sc ii} $\lambda 6355$ feature is redshifted by  $\unit[8000]{km \, s^{-1}}$ compared to that of the southernmost line of sight.  The velocity agreement is better for the lines of sight in the southern hemisphere, but the depths of the Si {\sc ii} $\lambda 5972$ and $\lambda 6355$ features do not match well.

\begin{figure}
  \centering
  \includegraphics[width=\columnwidth]{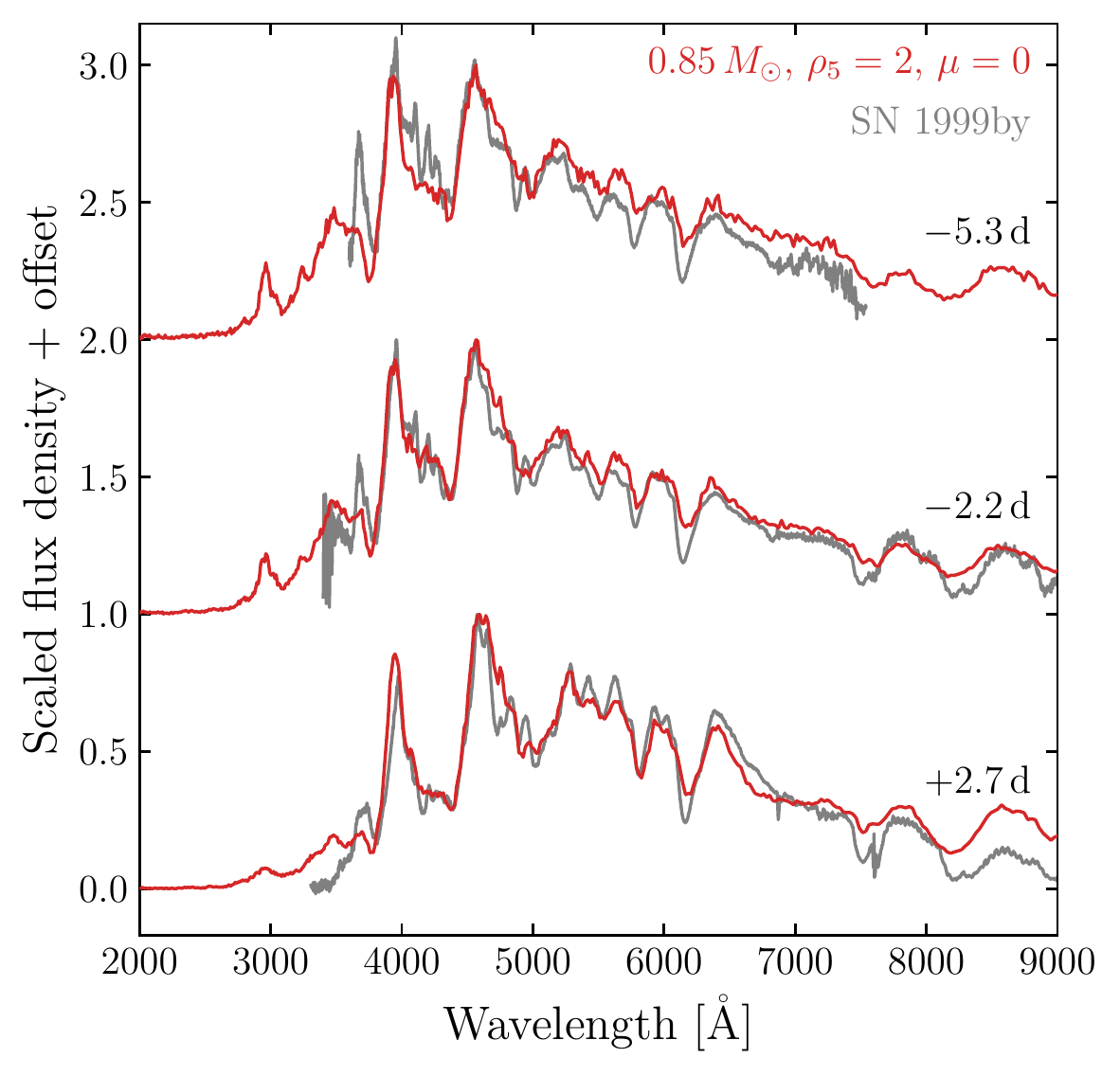}
  \caption{Spectra of the $0.85 \msol$, $\rho_5=2$ model viewed from the equator at the labeled times with respect to the time of $B$-band maximum.  Gray lines are spectra of SN~1999by at the same times, each of which is scaled by eye to provide a best fit to the theoretical spectrum.}
  \label{fig:spec_time_085}
\end{figure}

\begin{figure}
  \centering
  \includegraphics[width=\columnwidth]{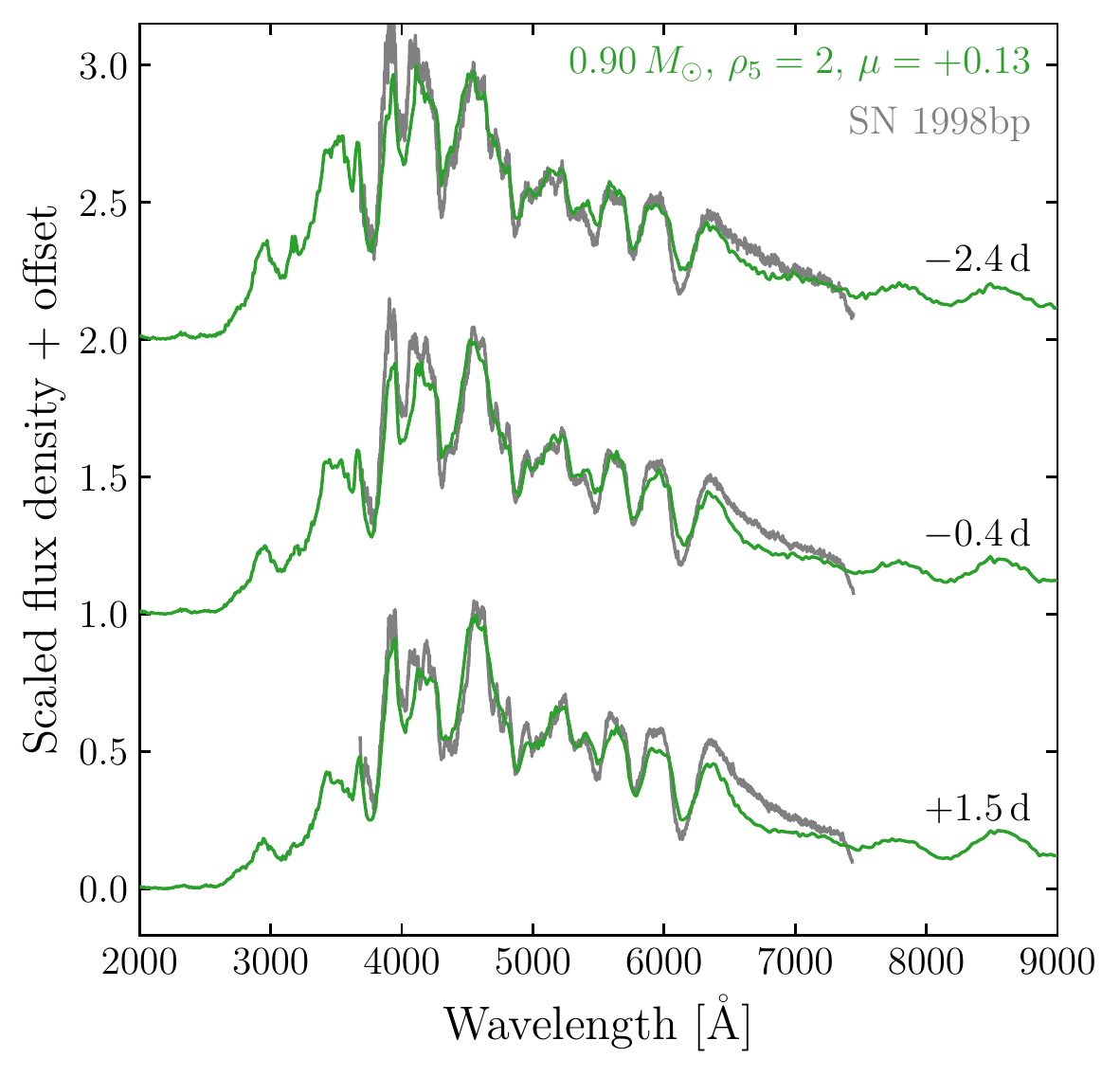}
  \caption{Same as Figure \ref{fig:spec_time_085}, but for the $0.90 \msol$, $\rho_5=2$ model compared to SN~1998bp.}
  \label{fig:spec_time_090}
\end{figure}

\begin{figure}
  \centering
  \includegraphics[width=\columnwidth]{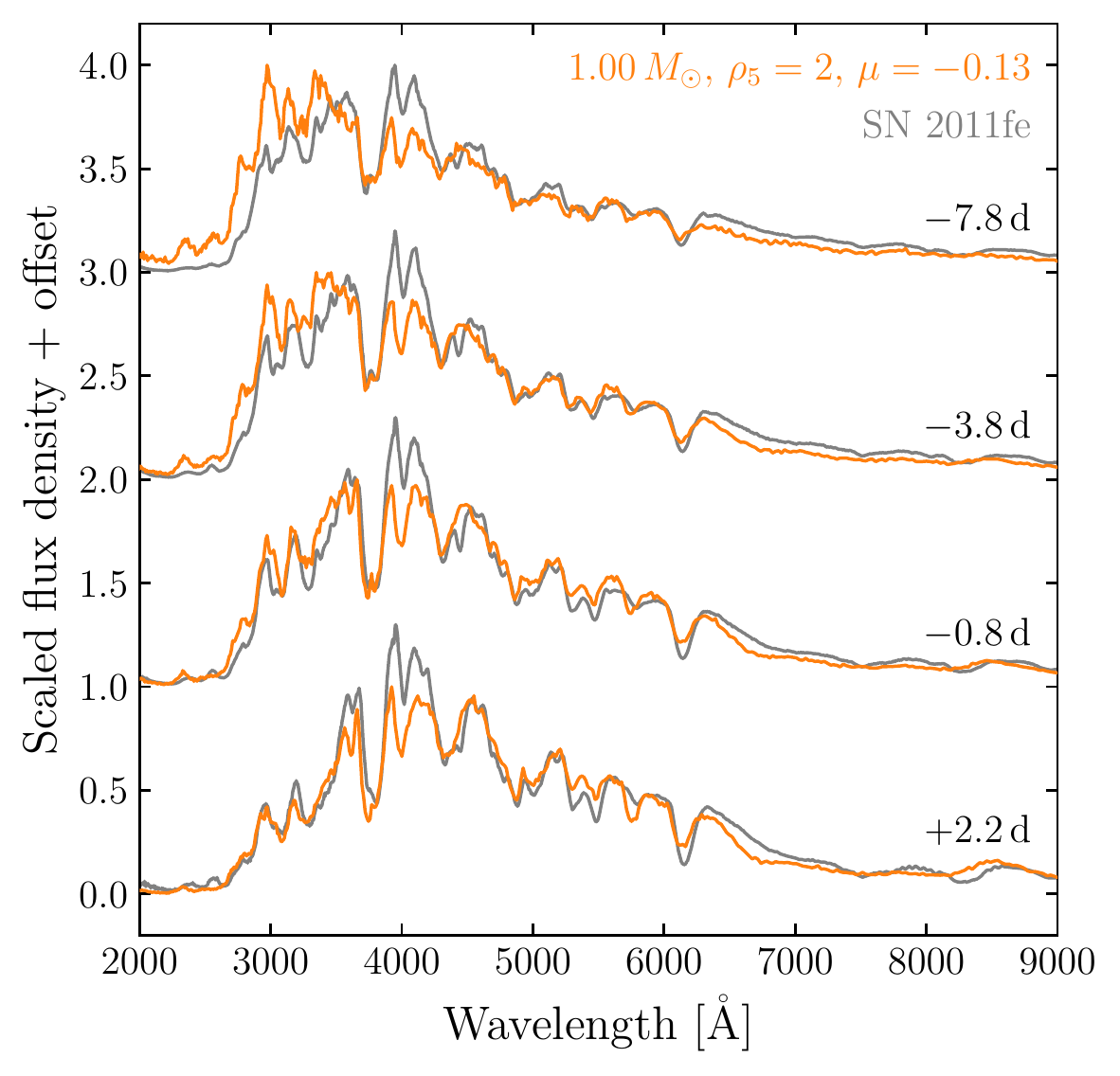}
  \caption{Same as Figure \ref{fig:spec_time_085}, but for the $1.00 \msol$, $\rho_5=2$ model compared to SN~2011fe.}
  \label{fig:spec_time_100}
\end{figure}

\begin{figure}
  \centering
  \includegraphics[width=\columnwidth]{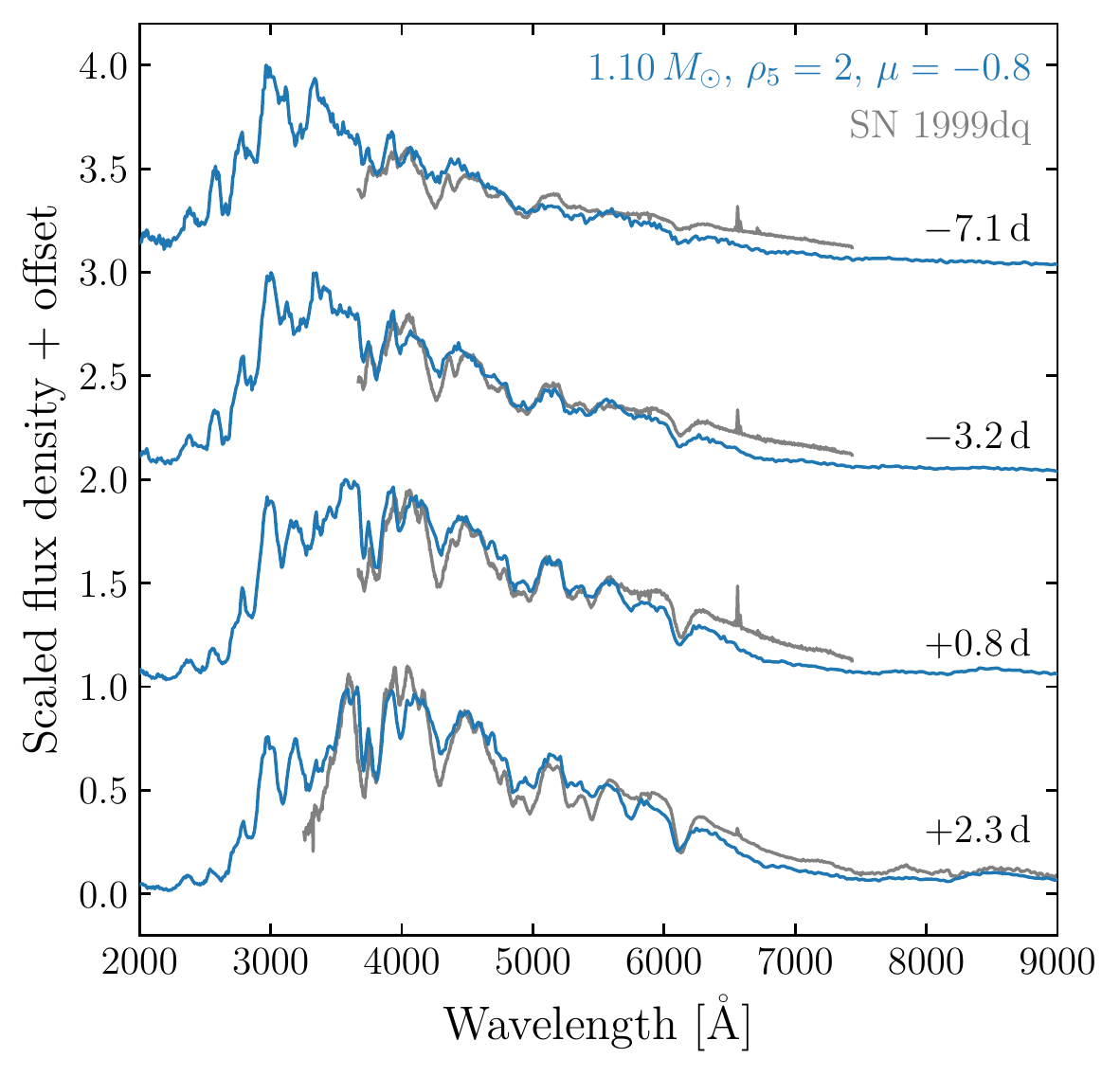}
  \caption{Same as Figure \ref{fig:spec_time_085}, but for the $1.10 \msol$, $\rho_5=2$ model compared to SN~1999dq.}
  \label{fig:spec_time_110}
\end{figure}

Figures \ref{fig:spec_time_085} -- \ref{fig:spec_time_110} show the spectral time evolution of each thin shell ($\rho_5=2$) model as viewed along the line of sight labeled in each figure.  Gray lines are observed spectra at the labeled times, each of which is arbitrarily scaled to provide a best fit by eye to the theoretical model.  The $0.85 \msol$ model is compared to SN~1999by \citep{garn04,math08a} in Figure \ref{fig:spec_time_085}.  Some aspects are well-reproduced (e.g., the Ti {\sc ii} trough deepens with time in a similar way for both the model and observed spectra), but some details do not match precisely, possibly as a consequence of our LTE, expansion opacity approximation \citep{ktn06,shen21a}: e.g., as discussed previously, the wavelength of the Si {\sc ii} $\lambda 6355$ absorption minimum is slightly discrepant, and the depths of some lines are not an exact match.  However, the overall correspondence is encouraging.
  
The time evolution of our $0.9 \msol$ model is shown in Figure \ref{fig:spec_time_090}, compared to that of SN~1998bp \citep{math08a}, which has been corrected for Milky Way reddening \citep{schl11a}.  The spectra do not cover as large a time window as in the other figures, but within the observed range, the model spectra evolve similarly to the observed spectra.  For example, the feature near $\unit[4300]{\AA}$ deepens and flattens with time for both the model and the observed spectra.  There are some slight discrepancies in the depths of some other absorption features, but there is good agreement overall.

Figure \ref{fig:spec_time_100} compares the evolution of the $1.00 \msol$ model to SN~2011fe \citep{mazz14a}.  Again, the theoretical model provides an adequate match to the observed spectrum in many respects.  However, as noted in the discussion of the maximum-light spectra (Figure \ref{fig:spec_mu_100_2}), the model spectra do not have as much absorption in the  Si {\sc ii} $\lambda 4130$ feature, and the absorption minima of the Si {\sc ii} $\lambda 4130$ and $\lambda 6355$ lines are somewhat blueshifted with respect to the observed spectra.

Our $1.10 \msol$ model is compared to SN~1999dq \citep{math08a,silv12b}, corrected for Milky Way reddening \citep{schl11a}, in Figure \ref{fig:spec_time_110}.  Unlike the previous three figures, for which lines of sight near the equator were used, this figures shows the evolution of a line of sight near the southern pole, for which the Si {\sc ii} $\lambda 6355$ velocity provides a better match to the observed velocity.  As with the previous comparisons, the match between the model and observed spectra is not exact, with discrepancies in the depths of several absorption features.  However, most of the absorption features across the optical range deepen with time in a similar way, and, in general, the agreement is satisfactory.


\section{Maximum light correlations}
\label{sec:maxcorr}

In this section, we perform an exploration of photometric and spectroscopic correlations in our model observables near maximum light.  We caution that, due to the LTE nature of our radiation transport calculations, there are almost certainly systematic offsets in the derived quantities, and so these  results should be viewed as merely suggestive.

\begin{figure*}
  \centering
  \includegraphics[width=\textwidth]{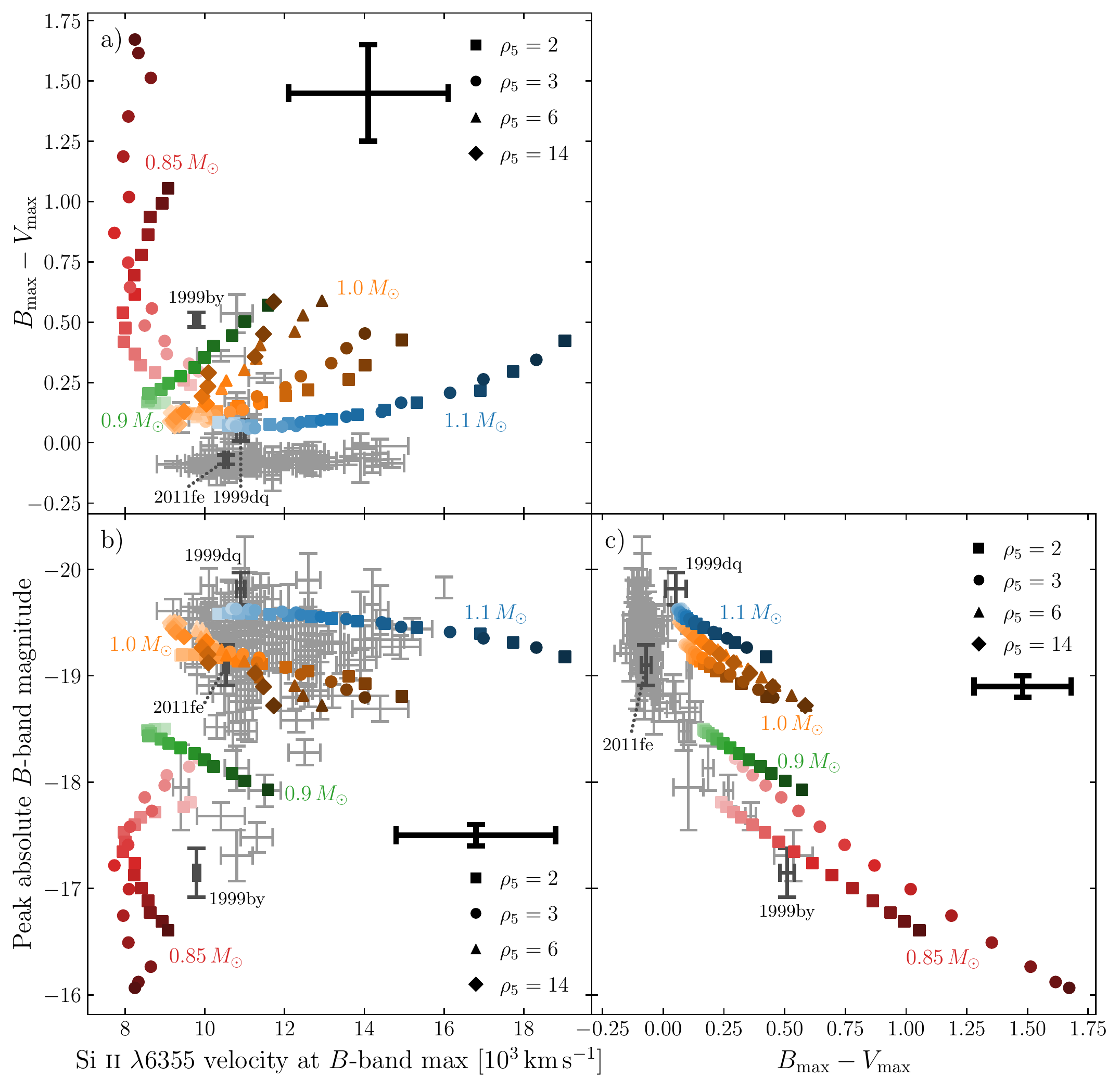}
  \caption{Correlations between a) $B_{\rm max}-V_{\rm max}$ and Si {\sc ii} $\lambda 6355$ velocity; b) $B_{\rm max}$ and Si {\sc ii}  velocity; and c)  $B_{\rm max}$ and $B_{\rm max}-V_{\rm max}$.  Colored points are our theoretical models, as labeled, with $\mu$ increasing with darkness.  Black error bars represent the maximum differences in the relevant  quantities between LTE and non-LTE results from \cite{shen21a}.  The velocity error bar includes an additional $\unit[1000]{km \, s^{-1}}$ of variance from the D$^6$ slingshot effect.  Gray error bars are observed SNe from \cite{burr20a}; dark gray error bars are individual SNe.}
  \label{fig:combined_maxcorr}
\end{figure*}

Figure \ref{fig:combined_maxcorr} compares three near-maximum light quantities: the peak absolute $B$-band magnitude ($B_{\rm max}$), the Si {\sc ii} $\lambda 6355$ velocity at the time of $B$-band maximum, as measured by the blueshift of the absorption minimum, and the color defined by the difference between $B_{\rm max}$ and the absolute $V$-band maximum magnitude, $V_{\rm max}$, which occurs several days after the time of $B$-band maximum.  Each panel shows a different comparison of two of the three quantities: color vs.\ Si velocity in panel a, $B_{\rm max}$ vs.\ Si velocity in panel b, and $B_{\rm max}$ vs.\ color in panel c.  In each panel, colored symbols represent results from our simulations as labeled, with $ \mu $ of the viewing angle increasing with darkness. 

Many similar comparisons have been made in observational data (e.g., \citealt{hach06a,fole11a,fole11b,blon12a,magu14a,zhen18a}).  Here, we use the recent compilation from \cite{burr20a}, which is shown as gray error bars in each panel.  Dark gray error bars represent SN~1999by \citep{garn04,hach06a}, SN~2011fe \citep{pere13a,silv15a,zhan16a}, and SN~1999dq \citep{stri06a,gane10a,zhen18a}.  Note that \cite{burr20a}'s compilation includes SNe from \cite{zhen18a}, who did not include color information, so some SNe in panel b are not represented in the other two panels.

The black error bar in each panel shows the maximal differences in the relevant quantities between the LTE and non-LTE results for the one-dimensional bare WD core detonations in \cite{shen21a}, as an estimate of the possible offset between the LTE results in this paper and future non-LTE calculations.  We also add an additional $\unit[1000]{km \, s^{-1}}$ of variance to the representative velocity error bars to account for the slingshot effect: if SNe~Ia arise from the ``dynamically driven double-degenerate double-detonation'' (D$^6$) scenario, and if the companion WD survives \citep{shen18b}, the SN ejecta will be born with a $\sim \unit[1000]{km \, s^{-1}}$ randomly oriented kick.  

We emphasize that the black error bars do not represent random, Gaussian errors in the theoretical results. Furthermore, there are likely systematic shifts in the relevant quantities between LTE and non-LTE results.  For example, the colors of the non-LTE simulations in \cite{shen21a} are consistently bluer than the LTE colors for WD masses $ \geq 0.9 \msol$, but redder for the lowest-mass $0.85 \msol$ models.  For the  highest Si {\sc ii} $\lambda 6355$ velocities, found in the $1.1 \msol$ explosion model, the non-LTE \texttt{CMFGEN} results are $\unit[1200]{km \, s^{-1}}$ slower than the LTE \texttt{Sedona} results, while for the $0.85 \msol$ model, the \texttt{CMFGEN} results are $\unit[300]{km \, s^{-1}}$ faster than the LTE \texttt{Sedona} results.

The $1.0$ and $1.1 \msol$ models satisfactorily cover the range of observed peak $B$-band magnitudes and Si {\sc ii} velocities for normal and overluminous SNe~Ia.  However, the $1.1 \msol$ models viewed from the northernmost lines of sight extend to higher velocities than observed in the \cite{burr20a} sample, reaching up to $ \unit[19000]{km \, s^{-1}}$.  The lack of observed SNe with such high velocities may be due to the relative rarity of $1.1 \msol$ explosions combined with the low probability of viewing such an explosion from the northernmost lines of sight.  Moreover, future non-LTE calculations may result in a systematic redshift of these highest velocities, as found by \cite{shen21a}.  On the other end of the mass range, the $0.85 \msol$ models yield Si {\sc ii} velocities that are $ \unit[1000-2000]{km \, s^{-1}}$ slower than observed in subluminous SNe~Ia.  The $0.85 \msol$ models' northernmost lines of sight also reach fainter peak $B$-band magnitudes than observed, by up to $\unit[1]{mag}$.

The largest discrepancy between our models and observed data is in the $B_{\rm max}-V_{\rm max}$ colors: the theoretical models are generally redder than observed SNe~Ia.  Focusing on the thin-shell $\rho_5=2$ and $\rho_5=3$ cases, the $1.0$ and $1.1 \msol$ models are up to $\unit[0.5]{mag}$ redder than observed SNe~Ia with the same peak $B$-band magnitudes or Si {\sc ii} $\lambda 6355$ velocities.  However, given the systematic blueward shift from LTE to non-LTE colors found by \cite{shen21a} for detonations of WDs with similar masses, this color discrepancy may disappear with more physical radiative transfer calculations.

Meanwhile, the $0.85 \msol$ models extend to much redder colors than observed, by as much as $\unit[1.1]{mag}$ for the $\rho_5=3$ model's northernmost line of sight.  These discrepancies in  color and peak $B$-band magnitude are far beyond the differences between \cite{shen21a}'s LTE and non-LTE results and constitute evidence that $0.85 \msol$ WDs do not explode as subluminous SNe~Ia.  There is a lower limit to the mass of a WD that can undergo a converging shock double detonation, because the critical ingoing shock strength necessary to trigger a core detonation increases with decreasing central density \citep{shen14a}; a $0.85 \msol$ WD may be below this mass limit.

It is also possible that $0.85 \msol$ WDs do explode but belong to other observed classes of transients.  Type Iax supernovae \citep{fole13a,jha17a} are one such class of candidates, but they possess even lower Si {\sc ii} $\lambda 6355$ velocities at $B$-band maximum,  $< \unit[7000]{km \, s^{-1}}$, than seen in our models.  Ca-strong transients \citep{pere10,kasl12,shen19a} are even fainter than the dimmest lines of sight of our $0.85 \msol$ models, with peak $B$-band magnitudes $>-15$.  

Another potential observational counterpart are the SN~2002es-like SNe, including SN~2006bt, PTF~10ops, and iPTF~14atg \citep{fole10b,magu11a,gane12,cao15a,whit15a,taub17a}.  These SNe form a fairly heterogeneous class, with Si {\sc ii} $\lambda 6355$ velocities at $B$-band maximum that range from $6000$ to $ \unit[10000]{km \, s^{-1}}$ and $B_{\rm max} $ ranging from $ -17.6$ to $-19$.  Unlike Type Iax supernovae, which strongly prefer young stellar populations, SN~2002es-like SNe explode preferentially in early-type galaxies, matching binary population synthesis predictions of the host galaxies of double WD mergers with primary masses $\sim 0.85 \msol$ \citep{shen17c}.  However, while iPTF~14atg's maximum light spectrum does provide a reasonable match to the $0.85 \msol$, $\rho_5=2$ model viewed from the equatorial line of sight, the large range of velocities observed for SN~2002es-likes and the lack of dimmer and very red SN~2002es-like SNe appears to rule out a correspondence between these models and this class as a whole.

In addition to the three maximum light observables we focus on in Figure \ref{fig:combined_maxcorr}, other spectral indicators have also been studied previously, including the time derivative of the Si {\sc ii} $\lambda 6355$ velocity, (pseudo-)equivalent widths and depths of various lines, and the velocity of nebular phase iron-group emission lines (e.g., \citealt{nuge95a,bene05a,bran06a,taub08a,maed10,maed10b,fole11b,blon11a,blon12a,livn20a}).  We choose to focus on $B_{\rm max}$, the Si {\sc ii} $\lambda 6355$ velocity, and $B_{\rm max}-V_{\rm max}$ because these three indicators are more robust to the effects of Monte Carlo noise and change less between LTE and non-LTE calculations than other features.  In particular, nebular phase spectra clearly necessitate non-LTE simulations, and the interpretation of these spectra is made even more difficult due to line blending and time variability of the inferred velocities \citep{blac16a,boty17a,grah17a}.  A more detailed examination of correlations among these and other model observables awaits future high-resolution, non-LTE radiative transfer calculations.


\section{Comparison to previous work}
\label{sec:prev}

Radiative transfer simulations of sub-Chandrasekhar-mass WD detonations have been carried out previously, but the majority of these have assumed spherical symmetry \citep{wtw86,ww94,sim10,wk11,blon17a,shen18a,shen21a,gold18a,poli19a,wygo19a,wygo19b,kush20a}.  While qualitative progress can be made in one dimension (see, e.g., the agreement with the \citealt{phil93a} relationship in the non-LTE, large nuclear reaction network study by \citealt{shen21a}), more detailed comparisons require multiple dimensions, due to the inherent asymmetry imparted by the laterally propagating helium shell detonation.

Another subset of radiation transport calculations have focused on explosions following double WD mergers in which the companion WD is completely disrupted \citep{pakm10,pakm11,pakm12b,pakm21a,moll14a,rask14a,vanr16a}.  These are qualitatively different from the simulations in this work, which assume that only one WD explodes, leaving the companion star, be it a WD or a non-degenerate star, mostly intact.  We thus do not attempt detailed comparisons to these studies.

The first multi-dimensional radiative transfer simulations of single sub-Chandrasekhar-mass WD detonations were performed by \cite{krom10}, with angle-averaged results reported in \cite{fink10}.  Due to possible issues with the implementation of detonation physics, the nucleosynthesis in the models used in these studies differs significantly from more recent work, making a direct comparison between our predicted observables difficult.  However, there is qualitative agreement between our study and that of \cite{krom10}.  Both sets of calculations find a strong dependence on viewing angle, with maximum light colors becoming bluer as the line of sight moves from the north pole, where the helium shell detonation is ignited, to the south pole.  Most models in both studies are also redder at maximum light than observed SNe with similar luminosities, although it is difficult to disentangle the effects of the radiative transfer approximations in both studies from the thicker helium shells present in the previous work.  The closest match in initial conditions for a quantitative comparison is between their model 2, initially consisting of a $0.92 \msol$ core and a $0.084 \msol$ helium-rich envelope, and our thickest shell $1.0 \msol$, $\rho_5=14$ model, which has a $0.90 \msol$ core and a $0.10 \msol$ helium-rich envelope.  Our model produces significantly more $^{56}$Ni ($0.56 \msol$) than theirs ($0.34 \msol$) and commensurately less intermediate-mass element material than in their model.  It is thus not surprising that \cite{krom10} find peak $B$-band magnitudes ranging from $-16.1$ to $-18.3$ depending on the line of sight, whereas we derive peak magnitudes of $-18.7$ to $-19.5$.  We note that \cite{sim12} also perform a similar study, but with even lower-mass WDs (total masses $ \leq 0.79 \msol$) and very thick helium shells of $0.21 \msol$ for which we do not have comparable models.

Using very similar methodology to \cite{boos21a}, \cite{town19a} simulated the two-dimensional detonation of a $1.0 + 0.02 \msol$ WD and approximated multi-dimensional radiative transfer by constructing one-dimensional spherical profiles from wedges of the two-dimensional explosion simulation.  The choice was made to renormalize the density profiles of these wedges to yield one-dimensional spheres with total masses equal to $1.0 \msol$, which is likely the reason for the different light curve shapes compared to the results in this work; in particular, the multi-band light curves viewed from the northern hemisphere evolve more rapidly in \cite{town19a}'s work.  However, the near-maximum-light spectra are similar, with Si {\sc ii} $\lambda 6355$ velocities in the southern hemisphere that match those of the present work and of SN~2011fe, while those in the northern hemisphere are similarly too fast by several thousand ${\rm km \, s^{-1}}$.

\cite{gron20a} performed several three-dimensional double detonation simulations.  The best comparisons between their models and ours are their M1a and M2a models and our $1.0 \msol$, $\rho_5=6$ model.  Their radiative transfer results are  much redder than ours, with angle-averaged values of $B_{\rm max}-V_{\rm max}= 0.9$ and $1.1$, while our angle-averaged results  yield $B_{\rm max}-V_{\rm max}= 0.2$.  However, the two sets of models do not differ drastically in terms of nucleosynthesis: total $^{56}$Ni yields differ by $<10\%$, with similar differences for other important elements.  Furthermore, while \cite{gron20a}'s M2a model is ignited via the ``scissors'' mechanism due to the enhanced carbon abundance in the helium shell, their M1a explodes in a similar way to our models, via a converging shock, so asymmetry is not the root cause of the discrepancies.

\cite{gron21a} update their previous results with a large suite of double detonation calculations.  However, they only present bolometric light curves; future work will provide more detailed radiative transfer analysis.  They find similar bolometric trends within each model: the hemisphere opposite the initial helium-shell detonation yields brighter lines of sight that fade more rapidly, at least in a bolometric sense.   There are slight quantitative differences when the observables are examined at a more detailed level (e.g., while their M10\_02 model provides a good match to our $1.0 \msol$, $\rho_5=3$ model, their M08\_03 model, which is close to our $0.85 \msol$, $\rho_5=2$ model, is somewhat brighter and fades more slowly), but the overall correspondence is fairly satisfactory.


\section{Conclusions}
\label{sec:conc}

In this work, we have performed multi-dimensional radiation transport calculations of the suite of sub-Chandrasekhar-mass WD double detonation models described in \cite{boos21a}.  We find broad agreement with the light curves and spectra of the entire range of non-peculiar SNe~Ia, from subluminous to overluminous examples.  Increasing the total mass of the exploding WD leads to brighter and bluer SNe and increases the velocities of absorption features in maximum-light spectra.  Varying the viewing angle from the southern hemisphere (where the carbon core detonation begins) to the northern hemisphere (where the helium detonation is ignited) yields dimmer and redder SNe and increases spectral velocities for most models.

There are several significant discrepancies between our theoretical models and observed SNe.  The Si {\sc ii} velocities at maximum light of the  $1.1 \msol$ models viewed from the northernmost lines of sight are at the limit of, and possibly faster than, the fastest velocities observed for SNe~Ia.  Meanwhile, the velocities of the $0.85 \msol$ models are slower than for any observed SNe~Ia, and the fluxes viewed from the north are fainter than for even the dimmest SNe~Ia.  Moreover, the maximum-light colors of all of our models are generally several tenths of a magnitude redder than for observed SNe, and as much as $\sim \unit[1]{mag}$ redder for the $0.85 \msol$ model's northern lines of sight.  The magnitude of the discrepancies for the $0.85 \msol$ models  may indicate that the minimum mass for a successful core detonation lies between $0.85$ and $0.90 \msol$.

Future calculations that include more accurate physics and initial conditions may alleviate the  color and velocity discrepancies for the higher masses.  The radiative transfer simulations in the present work assume LTE for the level and ionization state populations, but  \cite{shen21a} have found that non-LTE calculations yield corrections that will help to reduce the aforementioned differences.  Furthermore, the existence of the companion WD is neglected in this study.  If it is located in the northern hemisphere at the time of the explosion, the companion WD will alter the velocity and thermodynamic structure of the northernmost SN ejecta, possibly slowing it down and depositing enough energy to yield better color and velocity matches to observations.


\acknowledgments

We acknowledge helpful discussions with Alison Miller, and we thank Anthony Burrow for providing data and the anonymous referee for their review.  K.J.S., D.M.T., and S.B. received support for this work from NASA through the Astrophysics Theory Program (NNX17AG28G).  D.K.\ is supported in part by the U.S.\ Department of Energy, Office of Science, Office of Nuclear Physics, under contract number DE-AC02-05CH11231 and DE-SC0004658, and by a SciDAC award DE-SC0018297.  This research was supported in part by the Gordon and Betty Moore Foundation through grant GBMF5076, by a grant from the Simons Foundation (622817DK), and by the Exascale Computing Project (17-SC-20-SC), a collaborative effort of the U.S. Department of Energy Office of Science and the National Nuclear Security Administration.    This research used the Savio computational cluster resource provided by the Berkeley Research Computing program at the University of California, Berkeley (supported by the UC Berkeley Chancellor, Vice Chancellor for Research, and Chief Information Officer).  This research also used resources of the National Energy Research Scientific Computing Center (NERSC), a U.S.\ Department of Energy Office of Science User Facility located at Lawrence Berkeley National Laboratory, operated under Contract No.\ DE-AC02-05CH11231.


\software{\texttt{FLASH} \citep{fryx00,dube09a}, \texttt{matplotlib} \citep{hunt07a}, \texttt{MESA} \citep{paxt11,paxt13,paxt15a,paxt18a,paxt19a}, \sedona \citep{ktn06}}



\end{document}